\documentclass{article} % For LaTeX2e
\usepackage{nips13submit_e,times}
\usepackage{url}
\usepackage[ruled]{algorithm2e}
\usepackage{Definitions}
\usepackage{empheq}
\usepackage{graphicx}

\usepackage{times}

\newenvironment{tight_list}{
\begin{itemize}\vspace{-3mm}
\setlength{\itemsep}{1pt}
\setlength{\parskip}{0pt}
\setlength{\parsep}{0pt}
}{\end{itemize}\vspace{-3mm}}

\newcommand{\continmax}{\textsc{ConTinEst}\xspace}
\newcommand{\influmax}{\textsc{Influmax}\xspace}
\newcommand{\netrate}{\textsc{NetRate}\xspace}
\newcommand{\spm}{\textsc{SP1M}\xspace}
\newcommand{\pmia}{\textsc{PMIA}\xspace}

\title{Scalable Influence Estimation in \\Continuous-Time Diffusion Networks}
\author{
Nan Du$^{*}$ \qquad Le Song$^{*}$ \qquad Manuel Gomez-Rodriguez$^{\dagger}$
\qquad Hongyuan Zha$^{*}$\\
Georgia Institute of Technology$^{*}$ \qquad MPI for Intelligent
Systems$^{\dagger}$\\
\texttt{dunan@gatech.edu} \qquad \texttt{lsong@cc.gatech.edu} \\
\qquad \texttt{manuelgr@tue.mpg.de} \qquad \texttt{zha@cc.gatech.edu} \\
}

\nipsfinalcopy % Uncomment for camera-ready version

\begin{document}

\maketitle

\vspace{-5mm}

\begin{abstract}
If a piece of information is released from a media site, can we predict whether it may spread to one million web pages, in a month ?
% To answer this question, it is essential to develop diffusion models and estimation algorithms which can provide
% accurate and efficient estimates of the number of follow-ups, or the scope of influence, given a set of hypothetical source nodes and a time window of interest.
%This influence estimation problem is very challenging since both the time-sensitive nature of the problem and the issue of scalability need to be addressed simultaneously.
This influence estimation problem is very challenging since both the time-sensitive nature of the task and the requirement of scalability need to be addressed simultaneously.
%
% Previous work showed that continuous-time diffusion models provide significantly more accurate models for information, in terms of both recovering hidden diffusion structures from cascade data and predicting the
% timings of future events. However, subsequent uses of such models for influence estimation and maximization are severely limited by their scalability and inability to incorporate rich diffusion dynamics.
%
In this paper, we propose a randomized algorithm for influence estimation in continuous-time diffusion networks. Our algorithm can estimate the influence of
every node in a network with $|\Vcal|$ nodes and $|\Ecal|$ edges to an accuracy of $\epsilon$ using  $n=O(1/\epsilon^2)$ randomizations and up to logarithmic factors $O(n|\Ecal|+n|\Vcal|)$ computations. When used as a subroutine in a greedy influence maximization approach, our proposed algorithm is guaranteed to find a set of $C$ nodes with the influence of at least $(1 - 1/e)\operatorname{OPT} - 2C\epsilon$, where $\operatorname{OPT}$ is the optimal value. Experiments on both synthetic and real-world data show that the proposed algorithm can easily scale up to networks of millions of nodes while significantly improves over previous state-of-the-arts in terms of the accuracy of the estimated influence and the quality of the selected nodes in maximizing the influence.

\end{abstract}

\vspace{-5mm}
\section{Introduction}
\label{sec:introduction}
\vspace{-3mm}

\setlength{\abovedisplayskip}{4pt}
\setlength{\abovedisplayshortskip}{1pt}
\setlength{\belowdisplayskip}{4pt}
\setlength{\belowdisplayshortskip}{1pt}
\setlength{\jot}{3pt}

\setlength{\floatsep}{3ex}
\setlength{\textfloatsep}{3ex}

Motivated by applications in viral marketing~\cite{kleinberg_kdd03}, researchers have been studying the influence maximization problem: find a set of nodes whose initial adoptions of certain idea or product can trigger, \emph{in a time window}, the
largest expected number of follow-ups. For this purpose, it is essential to accurately and efficiently estimate the number of follow-ups of an arbitrary set of source nodes within the given time window. This is a challenging problem for that we need first accurately model the timing information in  cascade data and then design a scalable algorithm to deal with large real-world networks.
Most previous work in the literature tackled the influence estimation and maximization problems for infinite time window~\cite{kleinberg_kdd03, ChenWY09, ChenYZ10, goyal, LeskovecKGFVG07, Domingos02}.
However, in most cases, influence must be estimated or maximized up to a given time, \ie, a finite time window must be considered~\cite{influmax}. For example, a marketer would like to have her advertisement
viewed by a million people in one month, rather than in one hundred years. Such time-sensitive requirement renders those algorithms which only consider static information, such as network topologies, inappropriate
in this context.

A sequence of recent work has argued that modeling cascade data and information diffusion using \emph{continuous-time} diffusion networks can provide significantly more accurate models than discrete-time models~\cite{nan_nips2012, nan_aistats2013, DBLP:netrate, infopath, ZhoZhaSon13, ZhoZhaSon13b,manuel2013icml,ShuZha13}. There is a twofold rationale behind this modeling choice.
First, since follow-ups occur asynchronously, continuous variables seem more appropriate to represent them. Artificially discretizing the time axis into bins introduces additional tuning parameters, like the bin size,
which are not easy to choose optimally.
Second, discrete time models can only describe transmission times which obey an exponential density, and hence can be too restricted to capture the rich temporal dynamics in the data.
Extensive experimental comparisons on both synthetic and real world data showed that continuous-time models yield significant improvement in settings such as recovering hidden diffusion network structures from
cascade data~\cite{nan_nips2012, DBLP:netrate} and predicting the timings of future events~\cite{nan_aistats2013,manuel2013icml}.

However, estimating and maximizing influence based on continuous-time diffusion models also entail many challenges.
First, the influence estimation problem in this setting is a difficult graphical model inference problem, \ie, computing the marginal density of continuous variables in loopy graphical models. The exact answer can be
computed only for very special cases. For example, Gomez-Rodriguez et al.~\cite{influmax} have shown that the problem can be solved exactly when the transmission functions are exponential densities, by using
continuous time Markov processes theory. However, the computational complexity of such approach, in general, scales exponentially with the size and density of the network. Moreover, extending the approach
to deal with arbitrary transmission functions would require additional nontrivial approximations which would increase even more the computational complexity.
Second, it is unclear how to scale up influence estimation and maximization algorithms based on continuous-time diffusion models to millions of nodes. Especially in the maximization case, even a naive sampling
algorithm for approximate inference is not scalable: $n$ sampling rounds need to be carried out for each node to estimate the influence, which results in an overall $O(n |\Vcal| |\Ecal|)$ algorithm.
Thus, our goal is to design a scalable algorithm which can perform influence estimation and maximization in this regime of networks with millions of nodes.

In particular, we propose \continmax (\textbf{Con}tinous-\textbf{T}ime \textbf{In}fluence \textbf{Est}imation), a scalable ran\-do\-mized algorithm for influence estimation in a continuous-time diffusion network with
heterogeneous edge transmission functions. The key idea of the algorithm is to view the problem from the perspective of graphical model inference, and reduces the problem to a neighborhood estimation
problem in graphs.
Our algorithm can estimate the influence of
every node in a network with $|\Vcal|$ nodes and $|\Ecal|$ edges to an accuracy of $\epsilon$ using  $n=O(1/\epsilon^2)$ randomizations and up to logarithmic factors $O(n|\Ecal|+n|\Vcal|)$ computations. When used as a subroutine in a greedy influence maximization algorithm, our proposed algorithm is guaranteed to find a set of nodes with an influence of at least $(1 - 1/e)\operatorname{OPT} - 2C\epsilon$, where $\operatorname{OPT}$ is the optimal value.
Finally, we validate \continmax on both influence estimation and maximization problems over large synthetic and real world datasets. In terms of influence estimation, \continmax is much closer to the true influence and much faster than other
state-of-the-art methods. With respect to the influence maximization, \continmax allows us to find a set of sources with greater influence than other state-of-the-art methods.

\vspace{-3mm}
\section{Continuous-Time Diffusion Networks}
\label{sec:diffusionmodel}
\vspace{-3mm}

First, we revisit the continuous-time generative model for cascade data in social networks introduced in~\cite{DBLP:netrate}. The model associates each edge $j\rightarrow i$ with a transmission
function, $f_{ji}(\tau_{ji})$, a density over time, in contrast to previous discrete-time models which associate each edge with a fixed infection probability~\cite{kleinberg_kdd03}. Moreover, it also differs from discrete-time
models in the sense that events in a cascade are not generated iteratively in rounds, but event timings are sampled directly from the transmission function in the continuous-time model. 

{\bf Continuous-Time Independent Cascade Model.}
Given a \emph{directed} contact network, $\Gcal = (\Vcal,\Ecal)$, we use a continuous-time independent cascade model for modeling a diffusion process~\cite{DBLP:netrate}. The process begins with a
set of infected source nodes, $\Acal$, initially adopting certain \emph{contagion} (idea, meme or product) at time zero. The contagion is transmitted from the sources along their out-going edges to their
direct neighbors. Each transmission through an edge entails a \emph{random} spreading time, $\tau$, drawn from a density over time, $f_{ji}(\tau)$. We assume transmission times are independent
and possibly distributed differently across edges. Then, the infected neighbors transmit the contagion to their respective neighbors, and the process continues.
We assume that an infected node remains infected for the entire diffusion process. Thus, if a node $i$ is infected by multiple neighbors, only the neighbor that first infects node $i$ will be the \emph{true
parent}. As a result, although the contact network can be an arbitrary directed network, each cascade (a vector of event timing information from the spread of a contagion) induces a Directed Acyclic Graph (DAG).

{\bf Heterogeneous Transmission Functions.}
Formally, the transmission function $f_{ji}(t_i|t_j)$ for directed edge $j\rightarrow i$ is the conditional density of node $i$ getting infected at time $t_i$ given that node $j$ was infected at time $t_j$. We assume
it is shift invariant: $f_{ji}(t_i|t_j) = f_{ji}(\tau_{ji})$, where $\tau_{ji}:=t_i - t_j$, and nonnegative: $f_{ji}(\tau_{ji}) = 0$ if $\tau_{ji} < 0$. Both parametric transmission functions, such as the exponential and Rayleigh
function~\cite{DBLP:netrate,survival}, and nonparametric function~\cite{nan_nips2012} can be used and estimated from cascade data (see Appendix~\ref{app:transmission} for more details).

{\bf Shortest-Path property.} The independent cascade model has a useful property we will use later: given a sample of transmission times of all edges, the time $t_i$ taken to infect a node $i$ is the length of the
shortest path in $\Gcal$ from the sources to node $i$, where the edge weights correspond to the associated transmission times.

\vspace{-3mm}
\section{Graphical Model Perspectives for Continuous-Time Diffusion Networks}
\label{sec:graphicalmodel}
\vspace{-3mm}
The continuous-time independent cascade model is essentially a directed graphical model for a set of \emph{dependent} random variables, the infection times $t_i$ of the nodes, where the conditional independence
structure is supported on the contact network $\Gcal$ (see Appendix~\ref{app:graphicalmodel} for more details). More formally, the joint density of $\{t_i\}_{i \in \Vcal}$ can be expressed as
\begin{align}
	p\rbr{\{t_i\}_{i\in\Vcal}} = \prod\nolimits_{i \in \Vcal} p\rbr{t_i | \{t_j\}_{j \in \pi_i}},
	\label{eq:dag_factorization}
\end{align}
where $\pi_i$ denotes the set of parents of node $i$ in a cascade-induced DAG, and $p(t_i | \{t_j\}_{j \in \pi_i})$ is the conditional density of infection $t_i$ at node $i$ given the infection times of its parents.

Instead of directly modeling the infection times $t_i$, we can focus on the set of mutually \emph{independent} random transmission times $\tau_{ji} = t_i - t_j$. Interestingly, by switching from
a node-centric view to an edge-centric view, we obtain a fully factorized joint density of the set of transmission times
\begin{align}
  p\rbr{\{\tau_{ji}\}_{(j,i)\in \Ecal}} = \prod\nolimits_{(j,i)\in\Ecal} f_{ji}(\tau_{ji}),
\end{align}
Based on the Shortest-Path property of the independent cascade model, each variable $t_i$ can be viewed as a transformation from the collection of variables $\{\tau_{ji}\}_{(j,i)\in \Ecal}$.

More specifically, let $\Qcal_i$ be the collection of directed paths in $\Gcal$ from the source nodes to node
$i$, where each path $q\in \Qcal_i$ contains a sequence of directed edges $(j,l)$. Assuming all source nodes are infected at zero time, then we obtain variable $t_i$ via
\begin{align}
  t_i = g_i\rbr{\{\tau_{ji}\}_{(j,i)\in \Ecal}} := \min_{q \in \Qcal_i} \sum\nolimits_{(j,l)\in q} \tau_{jl},
\end{align}
where the transformation $g_i(\cdot)$ is the value of the shortest-path minimization. As a special case, we can now compute the probability of node $i$ infected before $T$ using a set of independent variables:
\begin{align}
  \Pr\cbr{t_i \leq T} = \Pr\cbr{g_i\rbr{\{\tau_{ji}\}_{(j,i)\in \Ecal}} \leq T}.
  \label{eq:equivalence}
\end{align}
%
% This will be very useful when we try to tackle the influence estimation problem discussed in the next section. \manuel{Several times you use, it will be useful in the future... but not now,
% produces a bit of impatient feeling}.
The significance of the relation is that it allows us to transform a problem involving a sequence of dependent variables $\{t_i\}_{i\in\Vcal}$ to one with independent variables $\{\tau_{ji}\}_{(j,i)\in\Ecal}$.
Furthermore, the two perspectives are connected via the shortest path algorithm in weighted directed graph, a standard well-studied operation in graph analysis.

\vspace{-3mm}
\section{Influence Estimation Problem in Continuous-Time Diffusion Networks}
\label{sec:influence}
\vspace{-3mm}

Intuitively, given a time window, the wider the spread of infection, the more influential the set of sources. We adopt the definition of influence as the average number of infected nodes
given a set of source nodes and a time window, as in previous work~\cite{influmax}. More formally, consider a set of $C$ source nodes $\Acal \subseteq \Vcal$ which gets infected at time zero, then, given a time window $T$, a node $i$ is infected
in the time window if $t_i\leq T$. The expected number of infected nodes (or the influence) given the set of transmission functions $\cbr{f_{ji}}_{(j,i)\in\Ecal}$ can be computed as
\begin{align}
  \sigma(\Acal,T)
  = \EE\left[\sum\nolimits_{i\in \Vcal}\II\cbr{t_i\leq T}\right]
  = \sum\nolimits_{i\in \Vcal} \EE\sbr{\II\cbr{t_i\leq T}}
  = \sum\nolimits_{i \in \Vcal} \Pr\cbr{t_i \leq T},
  \label{eq:influence}
\end{align}
where $\II\cbr{\cdot}$ is the indicator function %(taking value of $1$ if its argument is true and $0$ otherwise),
and the expectation is taken over the the set of \emph{dependent}
variables $\{t_i\}_{i\in\Vcal}$.
% Inspired by reinforcement learning, one can also study the discounted influence,
% $
% 	\sigma(\Acal,T,\eta) = \int_{0}^{T} \eta(t)\; \sigma(\Acal,t)\, dt,
% 	\label{eq:discounted_influence}
% $
% where $\eta(t)$ is an non-increasing discounting function (for instance, an exponential function $\eta(t) = r^t$ for some $0 < r < 1$). Intuitively, the earlier the nodes in $\Acal$ can infect other
% nodes the more influential they are. Compared with the measure in Eq.~\eq{eq:influence}, this discounted measure also captures the speed of infection and not just the final total infection number
% by time $T$. Importantly, we can recover Eq.~\eq{eq:influence} by setting $\eta(t) = \II\cbr{t=T}$.

Essentially, the influence estimation problem in Eq.~\eq{eq:influence} is an inference problem for graphical models, where the probability of event $t_i \leq T$ given sources in $\Acal$ can be obtained by summing out the possible configuration of other variables $\{t_j\}_{j\neq i}$. That is
\begin{align}
\label{eq:influence_computation}
\Pr\{t_i \leq T\} = \int_{0}^{\infty} \cdots \int_{t_i=0}^{T}\cdots \int_{0}^{\infty}	\rbr{\prod\nolimits_{j\in \Vcal} p\rbr{t_j | \{t_l\}_{l\in\pi_j}} } \rbr{\prod\nolimits_{j \in \Vcal} dt_j},
\end{align}
% \begin{align}
% \label{eq:influence_computation}
% 	\Pr\{t_i \leq T\} = \int_{0}^{\infty}
% 	\II\cbr{t_i \leq T} \rbr{\prod\nolimits_{j\in \Vcal} p\rbr{t_j | \{t_l\}_{l\in\pi_j}} } \rbr{\prod\nolimits_{j \in \Acal} \II\cbr{t_j = 0}} \prod\nolimits_{j \in \Vcal} dt_j,
% \end{align}
%
which is, in general, a very challenging problem. First, the corresponding directed graphical models can contain nodes with high in-degree and high out-degree. For example, in Twitter, a
user can follow dozens of other users, and another user can have hundreds of ``followees''. The tree-width corresponding to this directed graphical model can be very high,
and we need to perform integration for functions involving many continuous variables. Second, the integral in general can not be evaluated analytically for heterogeneous transmission functions, which
means that we need to resort to numerical integration by discretizing the domain $[0, \infty)$. If we use $N$ level of discretization for each variable, we would need to enumerate $O(N^{|\pi_i|})$ entries,
exponential in the number of parents.

Only in very special cases, can one derive the closed-form equation for computing $\Pr\{t_i \leq T\}$~\cite{influmax}. However, without further heuristic approximation, the computational complexity of the
algorithm is exponential in the size and density of the network.
%For instance,~\cite{influmax} proposed an approach for exponential transmission
%functions, where the special properties of exponential densitys is used to map the problem into a continuous time Markov process problem, and the computation can be carried out
%via a matrix exponential. However, without further heuristic approximation, the computational complexity of the algorithm is still exponential in the network density in general.
The intrinsic complexity of the problem entails the utilization of approximation algorithms, such as mean field algorithms or message passing algorithms.%and There are multiple ways to perform approximate inference in graphical models,
We will design an efficient randomized (or sampling) algorithm in the next section. %In comparison with other alternatives, our algorithm has a number of advantages, which we will also discuss in the next section.

\vspace{-3mm}
\section{Efficient Influence Estimation in Continuous-Time Diffusion Networks}
\label{sec:estimation}
\vspace{-3mm}

Our first key observation is that we can transform the influence estimation problem in Eq.~\eq{eq:influence} into a problem with \emph{independent} variables. Using relation in Eq.~\eq{eq:equivalence}, we have
\begin{align}
 \label{eq:key1}
  \sigma(\Acal,T)
  = \sum\nolimits_{i \in \Vcal} \Pr\cbr{g_i\rbr{\{\tau_{ji}\}_{(j,i)\in\Ecal}}\leq T}
  = \EE \sbr{\sum\nolimits_{i\in \Vcal} \II\cbr{g_i\rbr{\{\tau_{ji}\}_{(j,i)\in\Ecal}}\leq T}},
\end{align}
where the expectation is with respect to the set of independent variables $\{\tau_{ji}\}_{(j,i)\in\Ecal}$. This equivalent formulation suggests a naive sampling (NS) algorithm for approximating $\sigma(\Acal,T)$: draw
$n$ samples of $\{\tau_{ji}\}_{(j,i)\in\Ecal}$, run a shortest path algorithm for each sample, and finally average the results (see Appendix~\ref{app:ns} for more details). However, this naive sampling approach has a computational complexity of $O(n C |\Vcal||\Ecal| + n C |\Vcal|^2\log|\Vcal|)$ due to the repeated calling of the shortest path algorithm. This is quadratic to the network size, and hence not scalable to millions of nodes.

Our second key observation is that for each sample $\{\tau_{ji}\}_{(j,i)\in\Ecal}$, we are only interested in the neighborhood size of the source nodes, \ie, the summation $\sum_{i\in\Vcal} \II\cbr{\cdot}$ in
Eq.~\eq{eq:key1}, rather than in the individual shortest paths. Fortunately, the neighborhood size estimation problem has been studied in the theoretical computer science literature. Here, we adapt a
very efficient randomized algorithm by Cohen~\cite{cohen1997size} to our influence estimation problem.
This randomized algorithm has a computational complexity of $O(|\Ecal|\log|\Vcal|+|\Vcal|\log^2|\Vcal|)$ and it estimates the neighborhood sizes for \emph{all} possible single source node locations. Since it needs to
run once for each sample of $\{\tau_{ji}\}_{(j,i)\in\Ecal}$, we obtain an overall influence estimation algorithm with $O(n |\Ecal|\log|\Vcal|+n |\Vcal|\log^2|\Vcal|)$ computation, nearly linear in network size.
Next we will revisit Cohen's algorithm for neighborhood estimation.
% instead of $O(nC |\Vcal|^2 + n C |\Vcal||\Ecal|\log|\Ecal|)$.

\vspace{-3mm}
\subsection{Randomized Algorithm for Single-Source Neighborhood Size Estimation}
\vspace{-3mm}

Given a fixed set of edge transmission times $\{\tau_{ji}\}_{(j,i)\in\Ecal}$ and a source node $s$, infected at time $0$, the neighborhood $\Ncal(s,T)$ of a source node $s$ given a time window $T$ is the set of nodes within
distance $T$ from $s$,~\ie,~
\begin{align}
  \Ncal(s,T) = \cbr{i~\big|~g_i \rbr{\{\tau_{ji}\}_{(j,i)\in\Ecal}} \leq T,~i\in \Vcal}.
  \label{eq:neighbor}
\end{align}
Instead of estimating $\Ncal(s,T)$ directly, the algorithm will assign an exponentially distributed random label $r_i$ to each network node $i$. Then, it makes use of the fact that the minimum of a set of exponential random variables $\{r_i\}_{i \in \Ncal(s,T)}$ will also be a exponential random variable, but with its parameter equals to the number of variables. That is
if each $r_i \sim \exp(-r_i)$, then the smallest label within distance $T$ from source $s$, $r_\ast:=\min_{i\in \Ncal(s,T)} r_i$, will distribute as $r_\ast \sim \exp\cbr{-|\Ncal(s,T)| r_\ast}$. Suppose we randomize over the labeling $m$ times, and obtain $m$ such least labels, $\{r_\ast^u\}_{u=1}^m$. Then the neighborhood size can be estimated as
\begin{align}
%   |\Ncal(s,T)| \approx (m-1)/ (\sum\nolimits_{u=1}^m r_\ast^u).
%   \label{eq:estimate_size}
  |\Ncal(s,T)| \approx \frac{m-1}{\sum_{u=1}^m r_\ast^u}.
  \label{eq:estimate_size}
\end{align}
which is shown to be an unbiased estimator of $|\Ncal(s,T)|$~\cite{cohen1997size}.
This is an interesting relation since it allows us to transform the counting problem in~\eq{eq:neighbor} to a problem of finding the minimum random label $r_\ast$. The key question is whether we can compute the least label $r_\ast$ efficiently, given random labels $\{r_i\}_{i \in \Vcal}$ and any source node $s$.

% Furthermore, the above estimator is unbiased with variance $|\Ncal(s,T)|/(m-2)$~\cite{cohen1997size}.

Cohen~\cite{cohen1997size} designed a modified Dijkstra'{}s algorithm (Algorithm~\ref{a1}) to construct a data structure $r_\ast(s)$, called least label list, for each node $s$ to support such query. Essentially, the
algorithm starts with the node $i$ with the smallest label $r_i$, and then it traverses in breadth-first search fashion along the reverse direction of the graph edges to find all reachable nodes. For each reachable node
$s$, the distance $d_\ast$ between $i$ and $s$, and $r_i$ are added to the end of $r_\ast(s)$. Then the algorithm moves to the node $i'$ with the second smallest label $r_{i'}$, and similarly find all reachable nodes.
For each reachable node $s$, the algorithm will compare the current distance $d_\ast$ between $i'$ and $s$ with the last recorded distance in $r_\ast(s)$. If the current distance is smaller, then the current $d_\ast$
and $r_{i'}$ are added to the end of $r_\ast(s)$. Then the algorithm move to the node with the third smallest label and so on. The algorithm is summarized in Algorithm~\ref{a1} in Appendix~\ref{app:leastlabellist}.

Algorithm~\ref{a1} returns a list $r_\ast(s)$ per node $s\in \Vcal$, which contains information about distance to the smallest reachable labels from $s$. In particular, each list contains pairs of distance and random labels,
$(d,r)$, and these pairs are ordered as
\begin{align}
  \infty > \;&d_{(1)} > d_{(2)} > \ldots > d_{(|r_\ast(s)|)} = 0 \\
  &r_{(1)} < r_{(2)} < \ldots < r_{(|r_\ast(s)|)},
\end{align}
where $\{\cdot\}_{(l)}$ denotes the $l$-th element in the list. (see Appendix \ref{app:leastlabellist} for an example).
% (See Figure~\ref{demo} for an example).
If we want to query the smallest reachable random label $r_\ast$ for a given source $s$ and a time $T$, we only need to perform a binary search on the list for node $s$:
\begin{align}
  r_\ast = r_{(l)},~\text{where}~d_{(l-1)} > T \geq d_{(l)}.
\end{align}
Finally, to estimate $\abr{\Ncal(s,T)}$, we generate $m$~\iid~collections of random labels, run Algorithm~\ref{a1} on each collection, and obtain $m$ values $\cbr{r_\ast^u}_{u=1}^m$, which we use on Eq.~\eq{eq:estimate_size}
to estimate $|\Ncal(i,T)|$.

% \begin{figure}[t]
%   \vspace{-3mm}
%   \centering
%   \includegraphics[width=0.6\columnwidth]{f1.pdf}
%   {\small
%   \begin{tabular}{l}
%   $\bullet$ Node labeling : \\
%   ~~~~$r(e) < r(b) < r(d) < r(a) < r(c) < r(g) < r(f)$\\
%   $\bullet$ Neighborhoods: \\
% %   $\Ncal(c,0.5)=\{a,c\};~\Ncal(c,1)=\{a,b,c\};$\\
%   ~~~~$\Ncal(c,2)=\{a,b,c,e\};~\Ncal(c,3)=\{a,b,c,d,e,f\};$\\
% %   $\Ncal(c,4)=\{a,b,c,d,e,f,g\};$\\
%   $\bullet$ Least-label list: \\
%   ~~~~$r_\ast(c): (2,r(e)),(1,r(b)),(0.5,r(a)),(0,r(c))$ \\
%   $\bullet$ Query: $r_\ast(c,0.8) = r(a)$ \\
%   \end{tabular}
%   }\vspace{-3mm}
%   \caption{Graph $\Gcal=(\Vcal,\Ecal)$, edge weights $\{\tau_{ji}\}_{(j,i)\in\Ecal}$, and node labeling $\cbr{r_i}_{i\in\Vcal}$ with the associated output from Algorithm~\ref{a1}.}%\Note{thick line, larger arrow, larger numbers, specific node labels}}
% \label{demo}\vspace{-3mm}
% \end{figure}

The computational complexity of Algorithm~\ref{a1} is $O(|\Ecal|\log|\Vcal| + |\Vcal|\log^2|\Vcal|)$, with expected size of each $r_\ast(s)$ being $O(\log|\Vcal|)$. Then the expected time for querying $r_\ast$
is $O(\log\log|\Vcal|)$ using binary search. Since we need to generate $m$ set of random labels and run Algorithm~\ref{a1} $m$ times, the overall computational complexity for estimating the single-source
neighborhood size for all $s\in\Vcal$ is $O(m|\Ecal|\log|\Vcal| + m|\Vcal|\log^2|\Vcal| + m |\Vcal|\log\log|\Vcal|)$. For large scale network, and when $m\ll \min\{|\Vcal|,|\Ecal|\}$, this randomized algorithm can
be much more efficient than approaches based on directly calculating the shortest paths.
% \Note{the notation of $r_i$, $r_\ast(i)$ and $r_\ast(i,T)$ can be confusing.}

\vspace{-3mm}
\subsection{Constructing Estimation for Multiple-Source
Neighborhood Size}
\vspace{-3mm}

\label{sec:multisource}

When we have a set of sources, $\Acal$, its neighborhood is the union of the neighborhoods of its cons\-ti\-tuent sources
\begin{align}
  \Ncal(\Acal,T) = \bigcup\nolimits_{i \in \Acal} \Ncal(i,T).
\end{align}
This is true because each source independently infects its downstream nodes. Furthermore, to calculate the least label list $r_\ast$ corresponding to $\Ncal(\Acal,T)$, we can simply reuse the least label list $r_\ast(i)$
of each individual source $i \in \Acal$. More formally,
\begin{align}
  r_\ast = \min\nolimits_{i \in \Acal}~ \min\nolimits_{j \in \Ncal(i,T)} r_j,
\end{align}
where the inner minimization can be carried out by querying $r_\ast(i)$. Similarly, after we obtain $m$ samples of $r_\ast$, we can estimate $|\Ncal(\Acal,T)|$ using Eq.~\eq{eq:estimate_size}.
Importantly, very little additional work is needed when we want to calculate $r_\ast$ for a set of sources $\Acal$, and we can reuse work done for a single source. This is very different from a naive sampling approach where the sampling
process needs to be done completely anew if we increase the source set. In contrast, using the randomized algorithm, only an additional constant-time minimization over $|\Acal|$ numbers is needed.

\vspace{-3mm}
\subsection{Overall Algorithm}
\vspace{-3mm}

So far, we have achieved efficient neighborhood size estimation of $|\Ncal(\Acal, T)|$ with respect to a given set of transmission times $\{\tau_{ji}\}_{(j,i)\in\Ecal}$. Next, we will estimate the influence by averaging over multiple sets of samples for $\{\tau_{ji}\}_{(j,i)\in\Ecal}$. More specifically, the relation from~\eq{eq:key1}
\begin{align}
  \sigma(\Acal,T)
  = \EE_{\{\tau_{ji}\}_{(j,i)\in\Ecal}} \sbr{|\Ncal(\Acal,T)|}
  = \EE_{\{\tau_{ji}\}} \EE_{\{r^1,\ldots,r^m\}|\{\tau_{ji}\}}\sbr{\frac{m-1}{\sum_{u=1}^m r_\ast^u}},
\end{align}
% \begin{align}
%   \sigma(\Acal,T)
%   = \EE_{\{\tau_{ji}\}_{(j,i)\in\Ecal}} \sbr{|\Ncal(\Acal,T)|}
%   = \EE_{\{\tau_{ji}\}} \EE_{\{r^1,\ldots,r^m\}|\{\tau_{ji}\}}\sbr{\frac{m-1}{\sum_{u=1}^m r_\ast^u}},
% \end{align}
suggests the following overall algorithm

\fbox{
	\parbox{0.98\columnwidth}{
		Continuous-Time Influence Estimation (\continmax):\\
		\begin{tight_list}
			\item[1.] Sample $n$ sets of random transmission times
			$
				\{\tau_{ij}^l\}_{(j,i)\in\Ecal}~\sim~\prod\nolimits_{(j,i) \in \Ecal} f_{ji}(\tau_{ji})
			$
			\item[2.] Given a set of $\{\tau_{ij}^l\}_{(j,i)\in\Ecal}$, sample $m$ sets of random labels
			$
				\{r_i^u\}_{i\in \Vcal}~\sim~\prod\nolimits_{i \in \Vcal} \exp(-r_i)
			$
			\item[3.] Estimate $\sigma(\Acal,T)$ by sample averages
			$
				\sigma(\Acal,T)
				\approx \frac{1}{n} \sum_{l=1}^n \rbr{(m-1)/\sum_{u_l=1}^m r_\ast^{u_l}}
			$\\
		\end{tight_list}
	}
}

Importantly, the number of random labels, $m$, does not need to be very large. Since the estimator for $|\Ncal(A,T)|$ is unbiased~\cite{cohen1997size}, essentially the outer-loop of averaging over $n$ samples of random transmission times further reduces the variance of the estimator in a rate of $O(1/n)$. In practice, we can use a very small $m$ (\eg, $5$ or $10$) and still achieve good results, which is also confirmed by our later experiments. Compared to~\cite{ChenWY09}, the novel application of  Cohen's algorithm arises for estimating influence for multiple sources, which drastically reduces the computation by cleverly using the least-label list from single source. Moreover, we have the following theoretical guarantee (see Appendix~\ref{app:proof1} for proof)
\begin{theorem}
	\vspace{-2mm}
	Draw the following number of samples for the set of random transmission times
	\begin{align}
		n \geqslant \frac{C \Lambda}{\epsilon^2} \log\rbr{\frac{2 |\Vcal|}{\delta}}
	\end{align}
	where $\Lambda:= \max_{\Acal:\abr{\Acal}\leq C} 2\sigma(\Acal,T) / (m-2) + 2Var(\abr{\Ncal(\Acal,T)}) + 2 a \epsilon /3$ and $\abr{\Ncal(\Acal,T)}\leq a$, and for each set of random transmission times, draw $m$ set of random labels.	Then
	$
		\abr{\widehat{\sigma}(\Acal,T) - \sigma(\Acal,T)} \leqslant \epsilon
	$
	uniformly for all $\Acal$ with $\abr{\Acal}\leqslant C$, with probability at least $1 - \delta$.
	\vspace{-2mm}
\end{theorem}
The theorem indicates that the minimum number of samples, $n$, needed to achieve certain accuracy is related to the actual size of the influence $\sigma(\Acal, T)$, and the variance of the neighborhood size $|\Ncal(\Acal,T)|$ over the random draw of samples. The number of random labels, $m$, drawn in the inner loop of the algorithm will monotonically decrease the dependency of $n$ on $\sigma(\Acal, T)$. It suffices to draw a small number of random labels, as long as the value of $\sigma(\Acal,T) / (m-2)$ matches that of $Var(\abr{\Ncal(\Acal,T)})$. Another implication is that influence at larger time window $T$ is harder to estimate, since $\sigma(\Acal,T)$ will generally be larger and hence require more random labels.

\vspace{-3mm}
\section{Influence Maximization}
\vspace{-3mm}

Once we know how to estimate the influence $\sigma(\Acal, T)$ for any $\Acal \subseteq \Vcal$ and time window $T$ efficiently, we can use them in finding the optimal set of $C$ source nodes
$\Acal^* \subseteq \Vcal$ such that the expected number of infected nodes in $\Gcal$ is maximized at $T$. That is,~we seek to solve,
\begin{align}\label{maxinf}
\Acal^* = \argmax\nolimits_{|\Acal|\leqslant C}\;\sigma(\Acal, T),
\end{align}
where set $\Acal$ is the variable. The above optimization problem  is NP-hard in general.
%
%\begin{theorem}{\cite{influmax}}
%\label{np}
 %Given a network $\mathcal{G}=(\mathcal{V}, \mathcal{E})$, an observation window $T$, and a set of temporal dynamics $f_{ji}\in\mathcal{F}$, the heterogeneous influence maximization problem %defined by Equation~\ref{maxinf} is NP-hard.
%\end{theorem}
%
By construction, $\sigma(\mathcal{A},T)$ is a non-negative, monotonic nondecreasing function in the set of source nodes, and it can be shown that $\sigma(\Acal, T)$
satisfies a diminishing returns property called submodularity~\cite{influmax}.

A well-known approximation algorithm to maximize mo\-no\-to\-nic submodular functions is the \emph{greedy algorithm}. It adds nodes to the source node set $\Acal$ sequentially. In step $k$, it
adds the node $i$ which maximizes the \emph{marginal gain}
$
\sigma(\Acal_{k-1} \cup \{i\} ; T) - \sigma(\Acal_{k-1} ; T).
$
The greedy algorithm finds a source node set which achieves at least a constant fraction $(1-1/e)$ of the optimal~\cite{nemhauser1978analysis}. Moreover, lazy
evaluation~\cite{LeskovecKGFVG07} can be employed to reduce the required number of \emph{marginal gains} per iteration.
%
%Fortunately, \cite{FNGW} developed a greedy algorithm to find provably suboptimal solutions for the maximization of a monotonically nondecreasing submodular function:
%\begin{theorem}
%\cite{FNGW}  For non-negative and monotonic submodular function $f$, let $S$ be a set of size k obtained by selecting elements one at a time, each time choosing the element that provides
%the largest marginal increase in the function value. Then $f(S)\geqslant(1 - 1/e)\cdot f(S^{*})$, where $S^{*}$ maximizes $f$ over all $k$-element sets.
%\end{theorem}
%\cite{LeskovecKGFVG07} further
%proposed a Lazy Evaluation strategy by making use of the submodular property to significantly improve the speed of the greedy method, which is also used in our algorithm.
%
By using our influence estimation algorithm in each iteration of the greedy algorithm, we gain the following additional benefits:
%
% {\bf Parsimonious sampling.} %The evaluation of $\sigma(\Acal, T)$ is a cost step in the submodular optimization process.
% %
% $\sigma(\Acal, T)$ is the average of an unbiased estimator $\etheta$. Then, we can largely reduce the number $K$ of the inner samples to estimate $\etheta(W_i)$ for that the variance of
% each $\etheta(W_i)$ will be canceled out from each other in the final average, which significantly improves the speed without even a loss of accuracy.
%

First, at each iteration $k$, we do not need to rerun the full influence estimation algorithm (section~\ref{sec:multisource}). We just need to store the least label list $r_\ast(i)$ for each node $i\in \Vcal$ computed for a
single source, which requires expected storage size of $O(|\Vcal|\log|\Vcal|)$ overall.

Second, our influence estimation algorithm can be easily parallelized. Its two nested sampling loops can be parallelized in a straightforward way since the variables are independent of each other. However,
in practice, we use a small number of random labels, and $m \ll n$. Thus we only need to parallelize the sampling for the set of random transmission times $\cbr{\tau_{ji}}$. The storage of the least element lists can also be distributed.

However, by using our randomized algorithm for influence estimation, we also introduce a sampling error to the greedy algorithm due to the approximation of the influence $\sigma(\mathcal{A},T)$. Fortunately, the greedy algorithm is tolerant to such sampling noise, and a well-known result provides a guarantee for this case (following an argument in~\cite[Th.~7.9]{phdkrause2008}):
\begin{theorem}
\vspace{-2mm}
Suppose the influence $\sigma(\Acal, T)$ for all $\Acal$ with $\abr{\Acal}\leq C$ are estimated uniformly with error $\epsilon$ and confidence $1 - \delta$, the greedy algorithm returns a set of sources $\widehat{\Acal}$ such that
$\sigma(\widehat{\Acal},T) \geq (1-1/e) OPT - 2C\epsilon $ with probability at least $1-\delta$.
\vspace{-2mm}
\end{theorem}

\vspace{-3mm}
\section{Experiments}
\vspace{-3mm}
%
%We evaluate the accuracy of the estimated influence given by \continmax on several synthetic networks. Then, by incorporating \continmax into the greedy algorithm described above, we investigate the overall
%performance of influence maximization on several synthetic and real networks. We show that our approach significantly outperforms the state-of-the-art methods in terms of both speed and solution quality.
We evaluate the accuracy of the estimated influence given by \continmax and investigate the performance of influence maximization on synthetic and real networks. We show that our approach significantly outperforms the state-of-the-art methods in terms of both speed and solution quality.

{\bf Synthetic network generation.} We generate three types of  Kronecker
networks~\cite{LeskovecCKFG10}: (\emph{i}) core-periphery networks (parameter matrix: [0.9 0.5; 0.5 0.3]), which mimic the information diffusion traces in real world networks~\cite{DBLP:netinf},
(\emph{ii}) random networks ([0.5 0.5; 0.5 0.5]), typically used in physics and graph theory~\cite{easley2010}
and (\emph{iii}) hierarchical networks ([0.9 0.1; 0.1 0.9])~\cite{DBLP:netrate}. Next, we assign a pairwise transmission function for every directed edge in each type of network and set its parameters at random.
In our experiments, we use the Weibull distribution~\cite{survival},
$
f(t;\alpha, \beta)=\frac{\beta}{\alpha}\rbr{\frac{t}{\alpha}}^{\beta - 1}e^{-(t/\alpha)^{\beta}}, t\geqslant 0,
$
where $\alpha>0$ is a scale parameter and $\beta>0$ is a shape parameter.  The Weibull distribution (Wbl) has often been used to model lifetime events in survival analysis, providing more flexibility than an
exponential distribution~\cite{survival}.
We choose $\alpha$ and $\beta$ from 0 to 10 uniformly at random for each edge in order to have heterogeneous temporal dynamics. Finally, for each type of Kronecker network, we  generate 10 sample networks, each of which has different $\alpha$ and $\beta$ chosen for every edge.

\begin{figure}[t]
	\centering
	\renewcommand{\tabcolsep}{5pt}
	{\small
	\begin{tabular}{ccc}
	\includegraphics[width=0.25\textwidth, height=80pt]{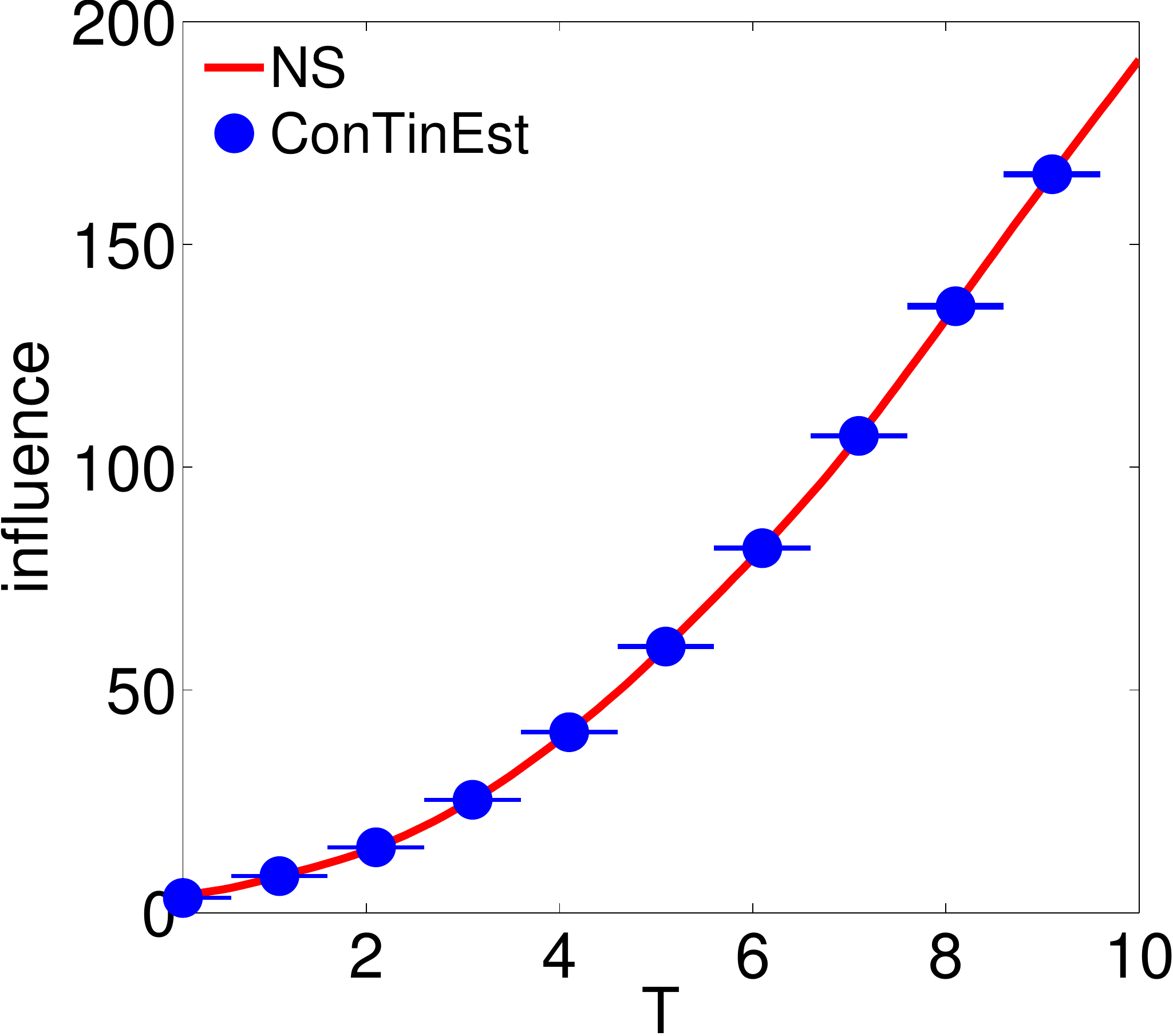} &
	\includegraphics[width=0.25\textwidth, height=80pt]{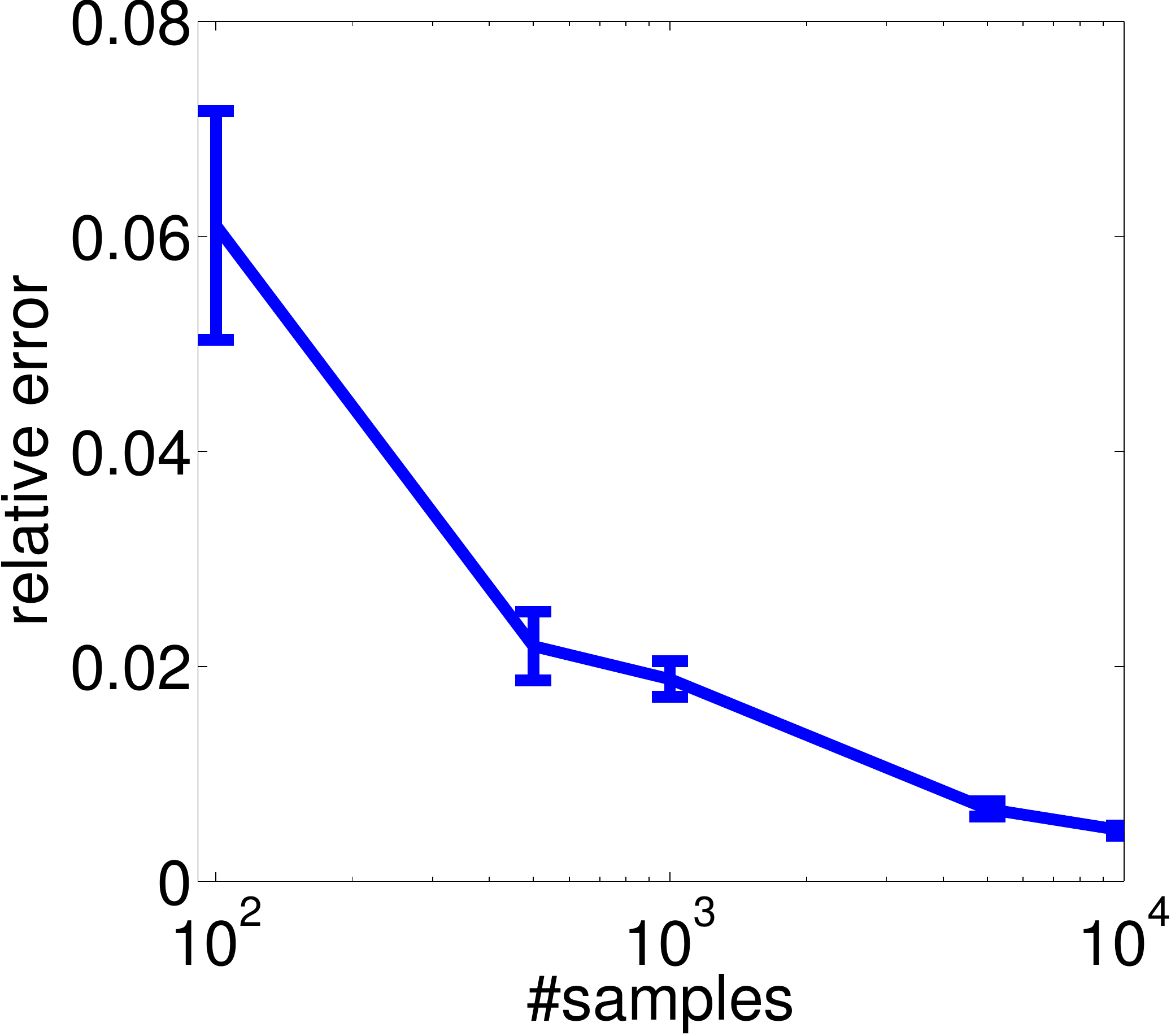} &
	\includegraphics[width=0.25\textwidth, height=85pt]{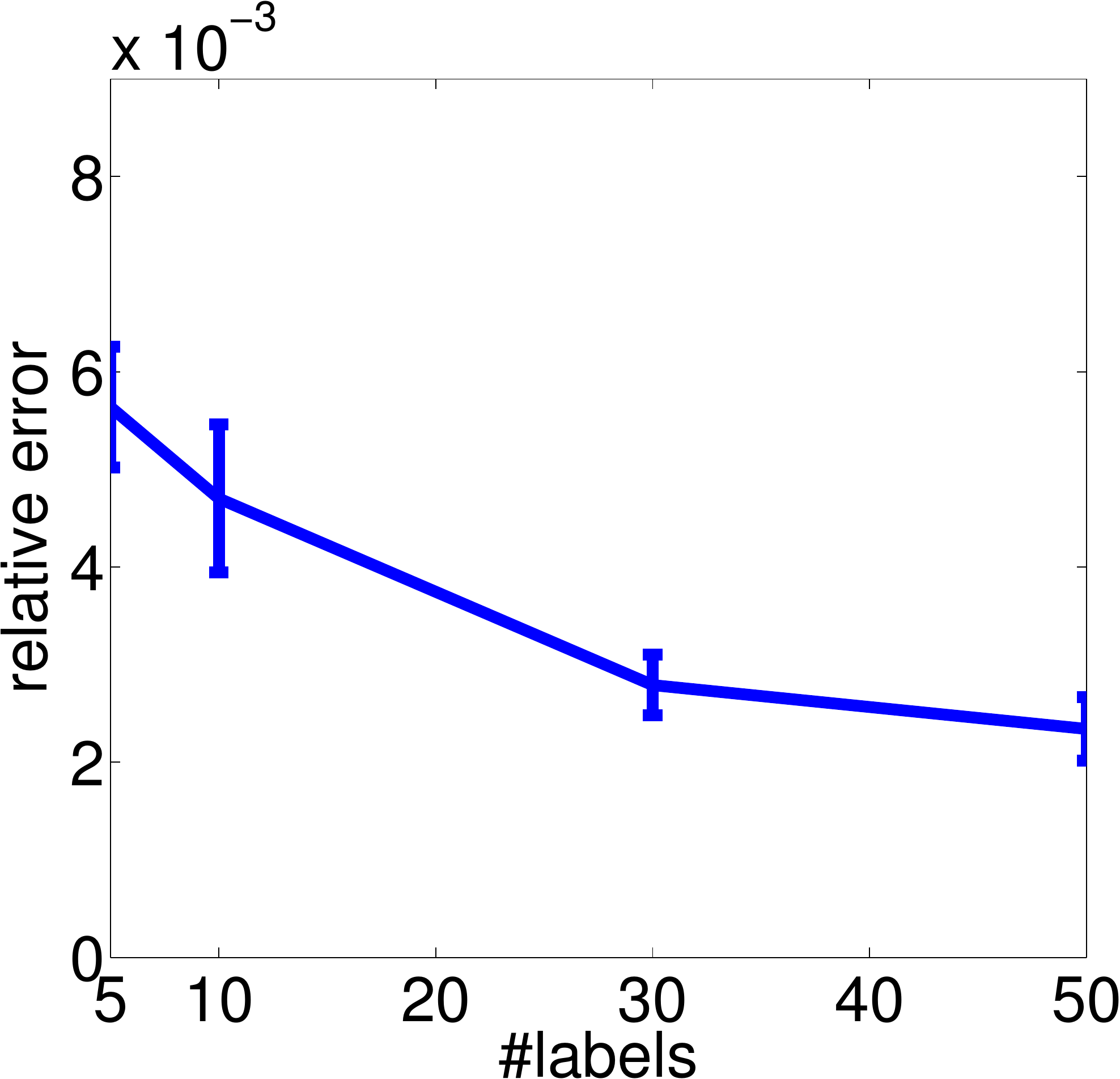} \\
	(a) Influence vs. time & (b) Error vs. \#samples & (c) Error vs. \#labels\\
	\end{tabular}
	}
	\vspace{-3mm}
 \caption{\small \label{accuracy_wbl} For core-periphery networks with 1,024 nodes and 2,048 edges,
 (a) estimated influence for increasing time window $T$, and (b) fixing $T=10$,
 relative error for increasing number of samples with 5 random labels, and (c) for
	increasing number of random labels with 10,000 random samples. }
	\vspace{-3mm}
\end{figure}

{\bf Accuracy of the estimated influence.} To the best of our knowledge, there is no analytical solution to the influence estimation given Weibull transmission function. Therefore, we compare \continmax with Naive Sampling (NS) approach (see Appendix~\ref{app:ns}) by considering the highest degree node in a network as the source, and draw 1,000,000 samples for NS to obtain near ground truth.
Figures~\ref{accuracy_wbl}(a) compares \continmax with the ground truth provided by NS at different time window $T$, from $0.1$ to $10$ in corre-periphery networks. For \continmax, we generate up to 10,000 random samples (or
set of random waiting times), and 5 random labels in the inner loop. In all three networks, estimation provided by \continmax fits accurately the ground truth, and the relative error decreases quickly as we increase the number of samples and labels (Figures~\ref{accuracy_wbl}(b) and~\ref{accuracy_wbl}(c)). For 10,000 random samples with 5 random labels, the relative error is smaller than 0.01. (see Appendix~\ref{app:exp} for additional results on the random and hierarchal networks)

\begin{figure}[t]
	\centering
  \renewcommand{\tabcolsep}{0pt}
	{\small
	\begin{tabular}{ccc}
		\includegraphics[width=0.25\textwidth, height = 80pt]{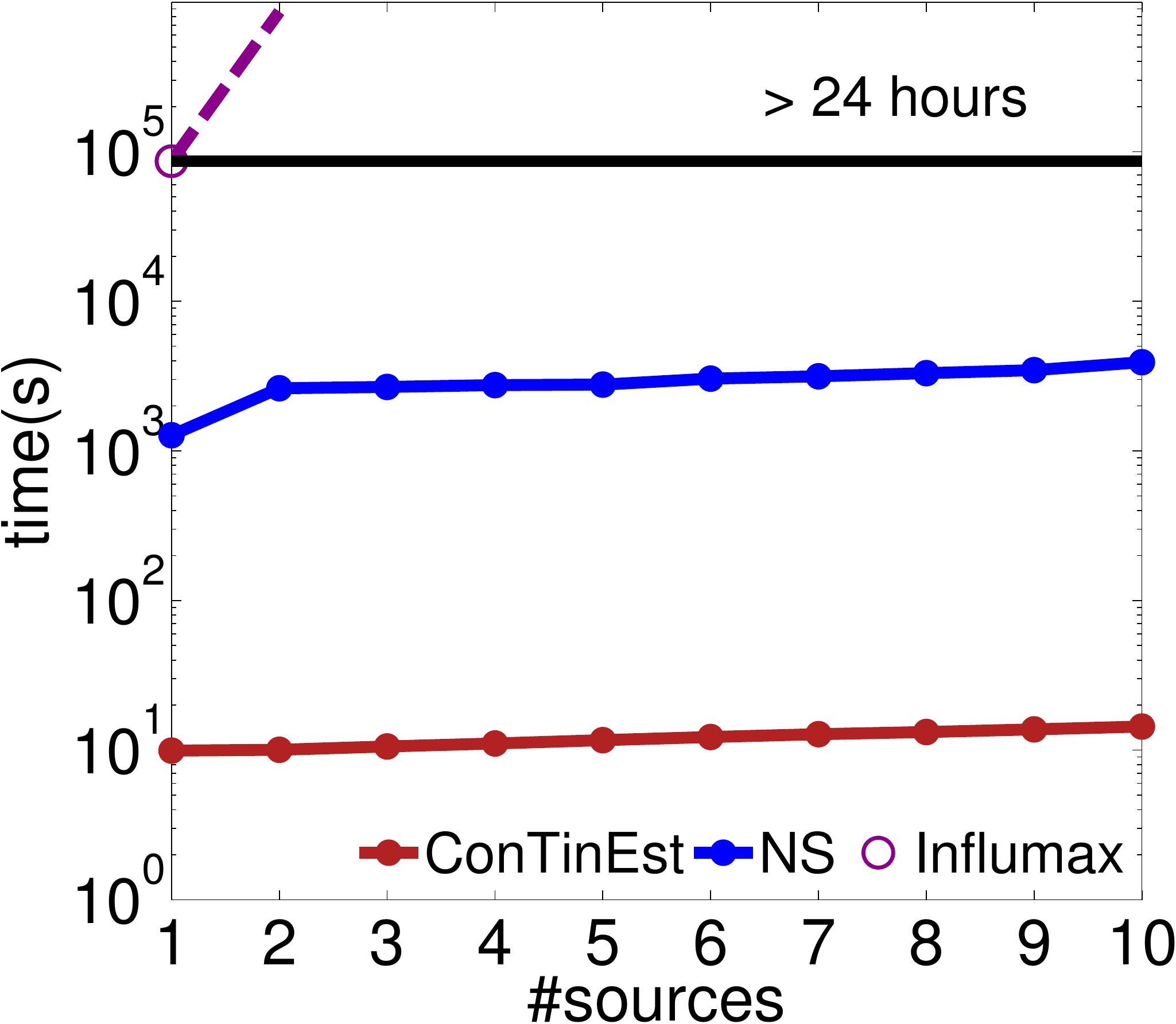} &
		\includegraphics[width=0.25\textwidth, height = 80pt]{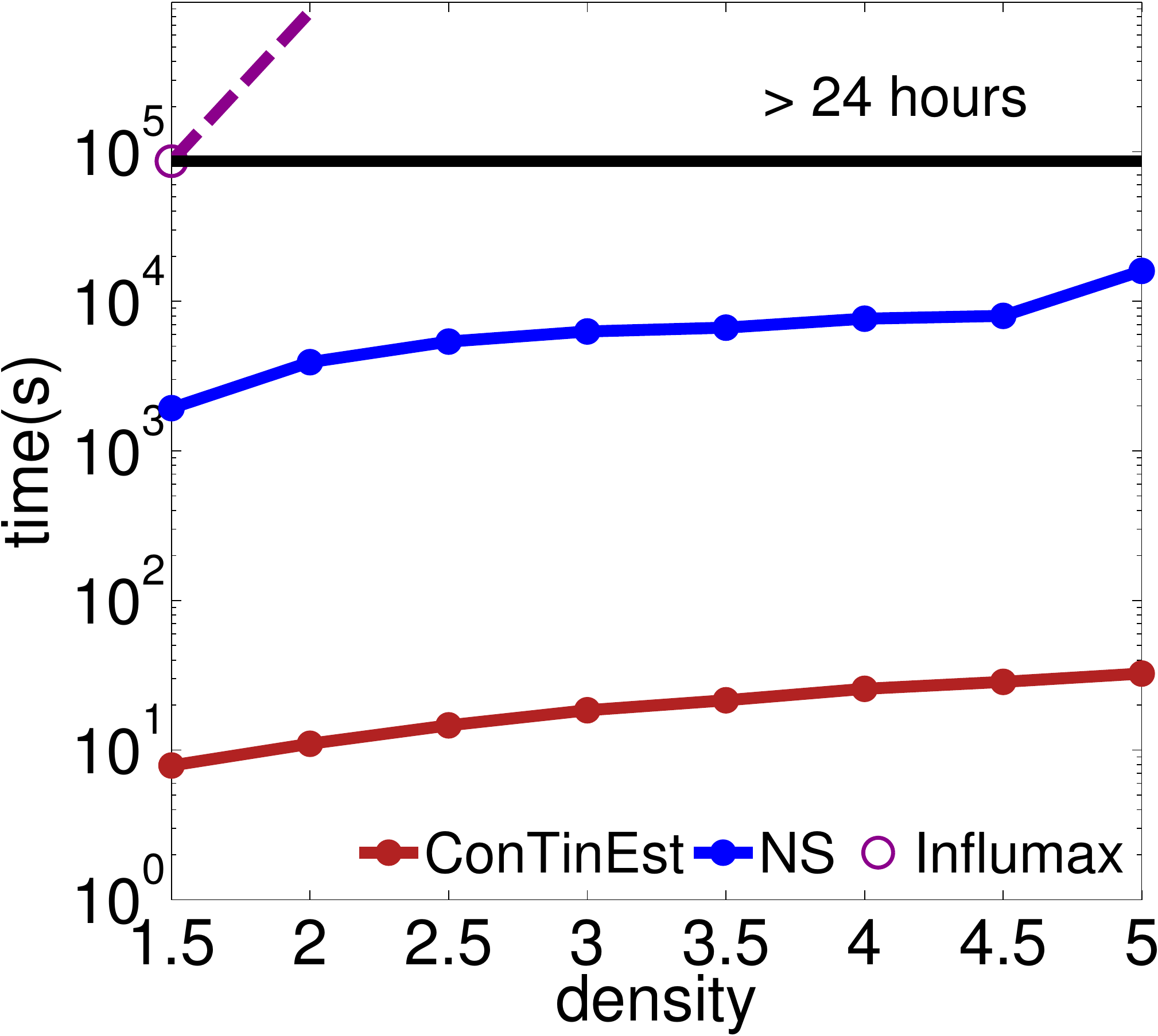} &
		\includegraphics[width=0.25\textwidth, height = 80pt]{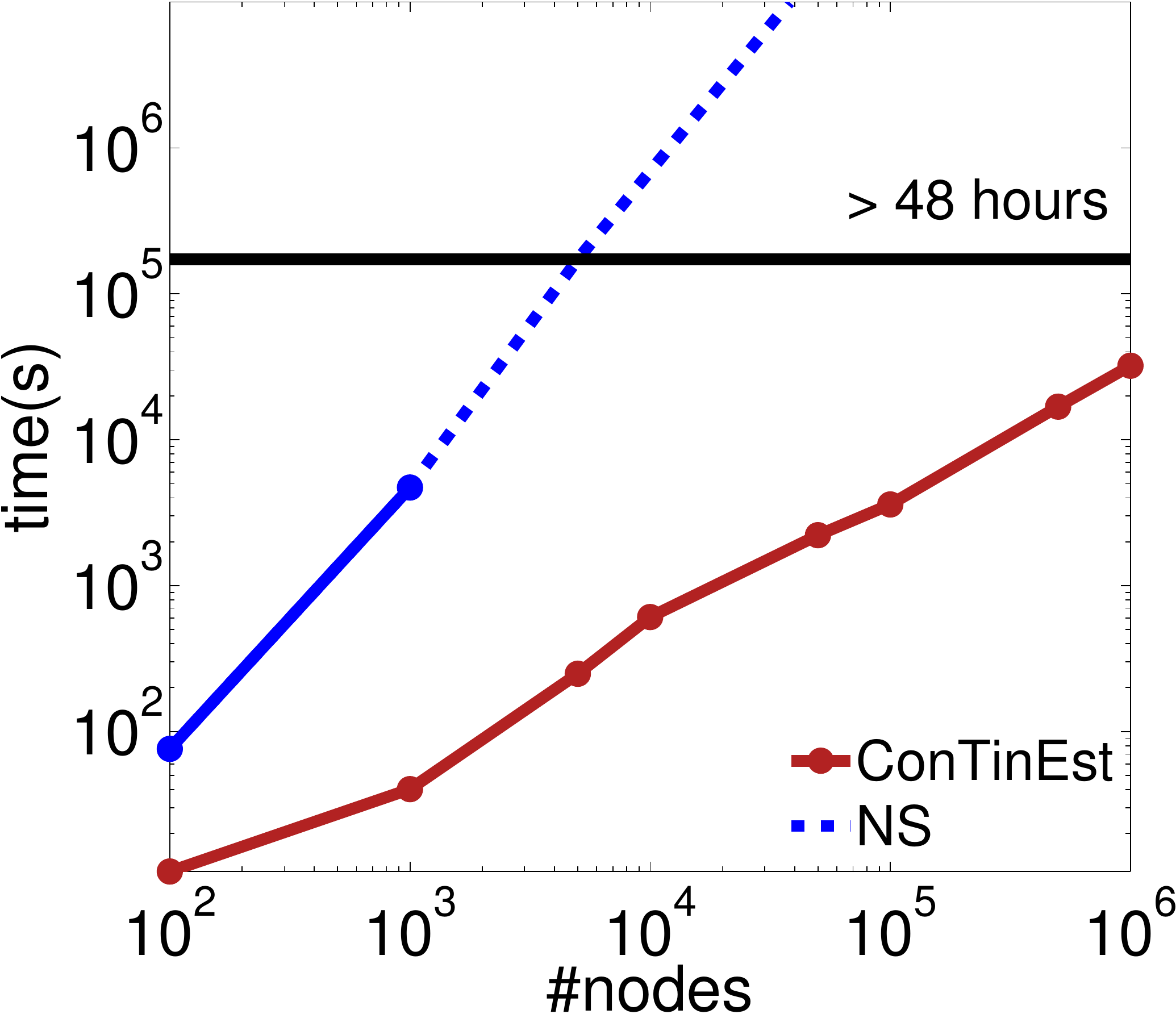} \\
		(a) Run time vs. \# sources & (b) Run time vs. network density &  (c)  Run time vs. \#nodes
	\end{tabular}
	}
	\vspace{-3mm}
	\caption{\small \label{density_speed} For core-periphery networks with $T = 10$, runtime for (a) selecting increasing number of sources in networks of 128 nodes and 320 edges; for  (b)selecting 10 sources in networks of 128 nodes with increasing density; and for (c) selecting 10 sources with increasing network size from 100 to 1,000,000 fixing 1.5 density.}
	\vspace{-3mm}
\end{figure}

{\bf Scalability.} We compare \continmax to previous state-of-the-art \influmax~\cite{influmax} and the Naive Sampling (NS) method in terms of run time for the continuous-time influence estimation and maximization. For \continmax, we draw 10,000 samples in the outer loop, each having 5 random labels in the inner loop. For NS, we also draw 10,000 samples. The first two experiments are carried out in a single 2.4GHz processor.
First, we compare the performance for increasing number of selected sources (from 1 to 10) by fixing the core-periphery networks to 128 node network and 320 edges and time window to 10 (Figure~\ref{density_speed}(a)). When the number of selected sources is 1, different algorithms essentially spend time estimating the influence for each node. \continmax outperforms
other methods by order of magnitude and for the number of sources larger than 1, it can efficiently reuse computations for estimating influence for individual nodes. Dashed lines mean that a method did not finish in 24 hours, and the estimated run time is plotted.
Next, we compare the run time for selecting 10 sources on core-periphery networks of 128 nodes with increasing densities (or the number of edges) (Figure~\ref{density_speed}(a)).
Again, \influmax and NS are order of magnitude slower due to their respective exponential and quadratic computational complexity in network density. In contrast, the run time of \continmax only increases slightly with the increasing density since its computational complexity is linear in the number of edges (see Appendix~\ref{app:exp} for additional results on the random and hierarchal networks).
Finally, we evaluate the speed on large core-periphery networks, ranging from 100 to 1,000,000 nodes with density 1.5 in Figure~\ref{density_speed}(c). We report the parallel run
time only for \continmax and NS (both are implemented by MPI running on 192 cores of 2.4Ghz) since \influmax is not scalable.
In contrast to NS, the performance of \continmax increases linearly with the network size and can easily scale up to one million nodes.

\begin{figure}[t]
% \vspace{-3mm}
\hspace{-4mm}
	\centering
	\renewcommand{\tabcolsep}{0pt}
	{\small
	\begin{tabular}{ccc}
		\includegraphics[width=0.25\columnwidth, height = 80pt]{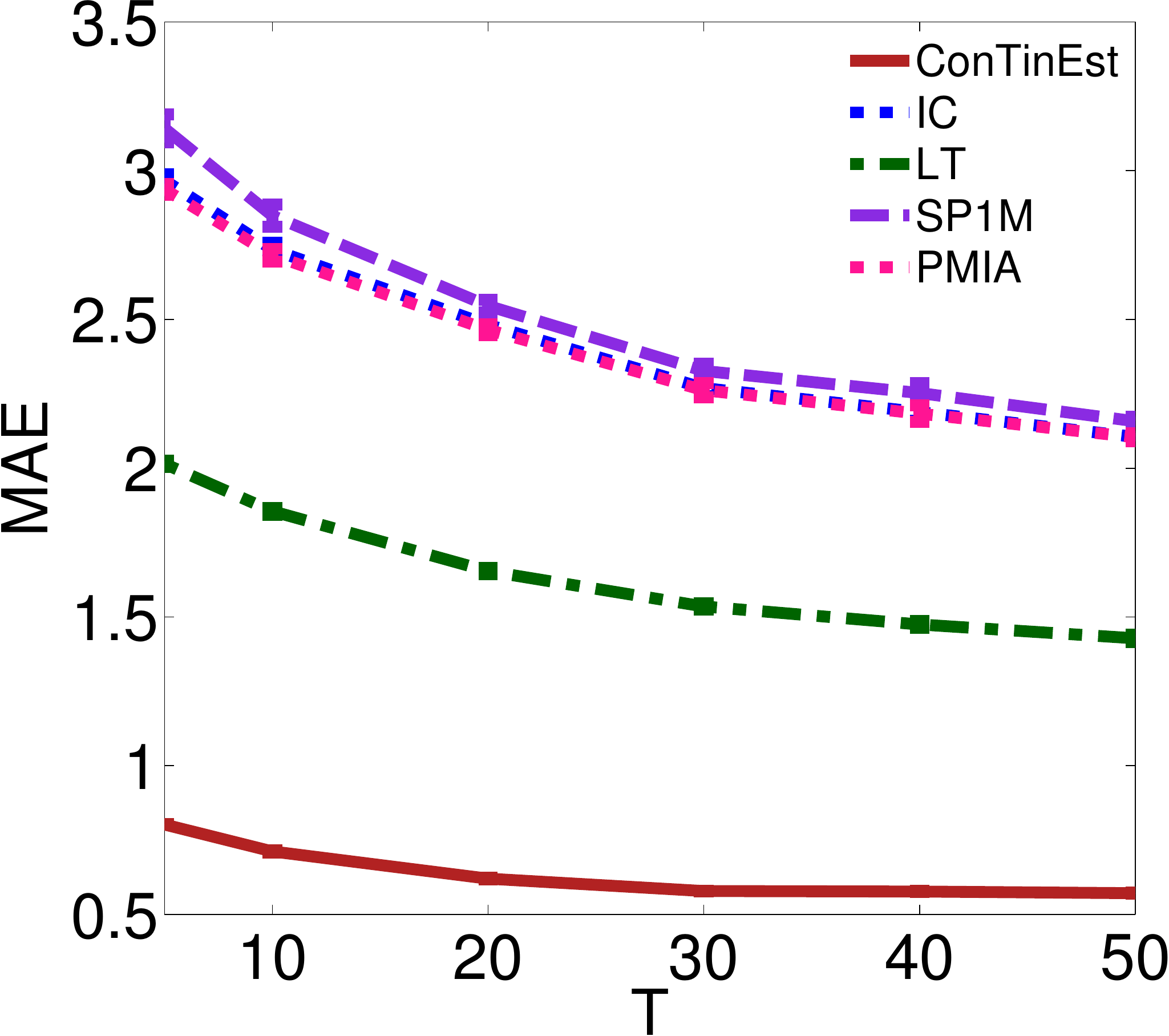} &
		~\includegraphics[width=0.25\columnwidth, height = 80pt]{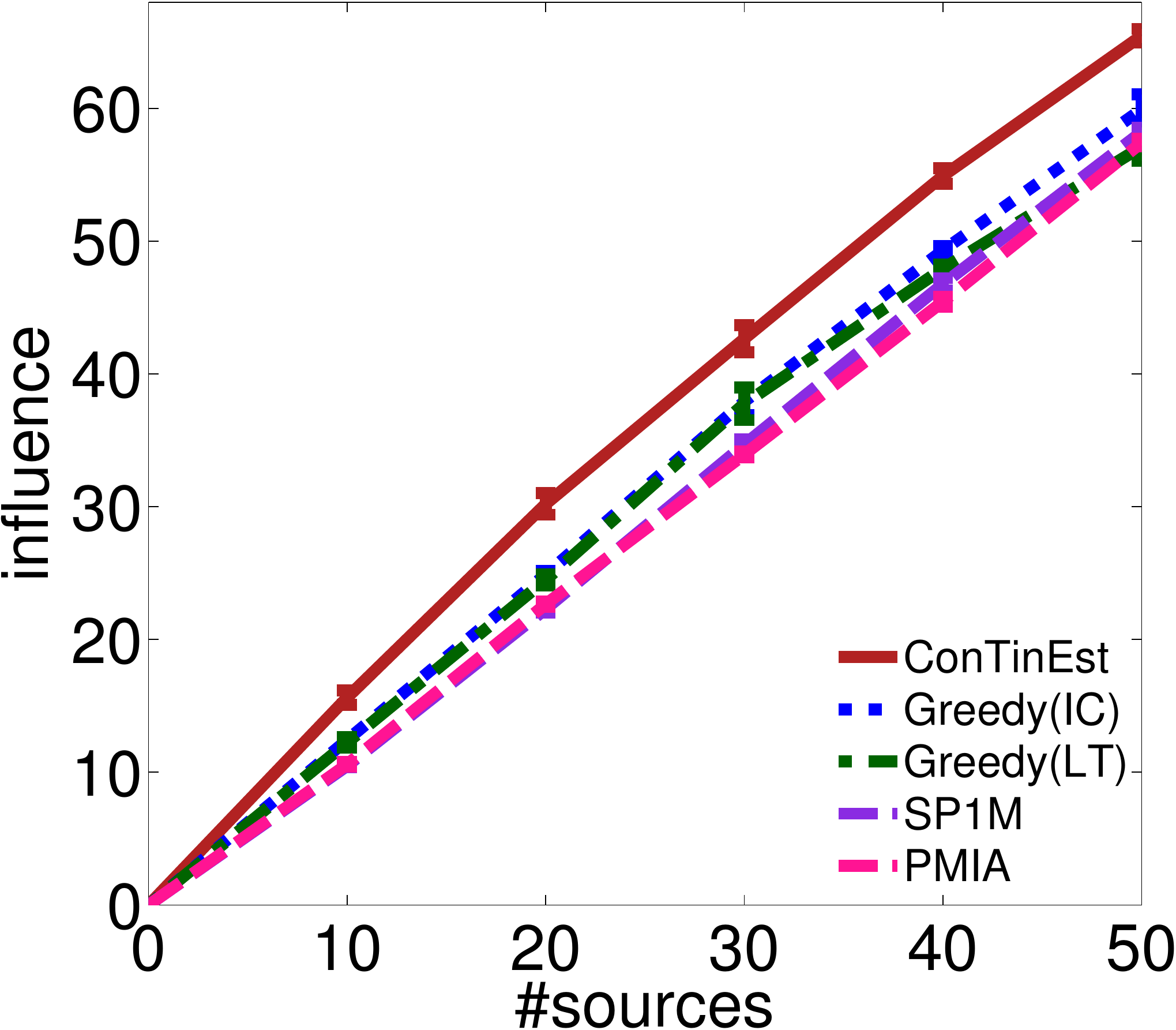} &
		~\includegraphics[width=0.25\columnwidth, height = 80pt]{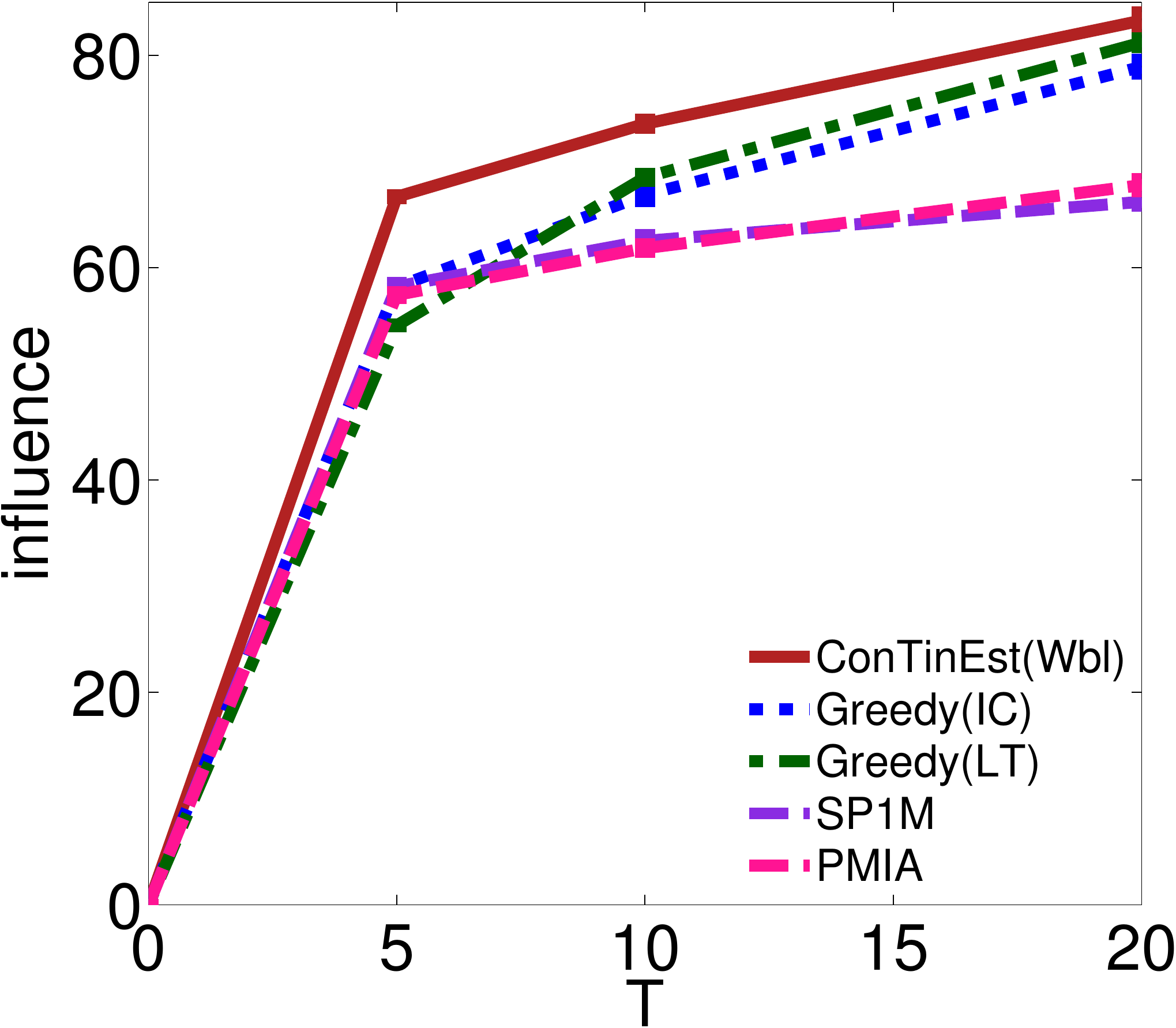} \\
	(a) Influence estimation error& (b) Influence vs. \#sources  & (c) Influence vs. time\\
	\end{tabular}
	}
	\vspace{-3mm}
	\caption{\small \label{real_influence} In MemeTracker dataset, (a) comparison of the accuracy of the estimated influence in terms of mean absolute error, (b) comparison of the influence of the selected nodes by fixing the observation window $T=5$ and varying the number sources, and (c) comparison of the influence of the selected nodes by by fixing the number of sources
	to 50 and varying the time window.}
	\vspace{-3mm}
 \end{figure}

{\bf Real-world data.}
We first quantify how well each method can estimate the true influence in a real-world dataset. Then, we evaluate the solution quality of the selected sources for influence maximization.
We use the MemeTracker dataset~\cite{DBLP:conf/kdd/LeskovecBK09} which has 10,967 hyperlink cascades among 600 media sites.
%each of which time-stamps the creation of articles referring to each other about similar pieces of information.
%We use MemeTracker dataset~\cite{DBLP:conf/kdd/LeskovecBK09}, where we trace the flow of information from one site to another by using the hyperlinks between articles and posts.
%We have extracted around 100,000 hyperlink cascades, each of which is a collection of time-stamps which record the creation of posts and articles that refer to each other about similar pieces of information.
%
%A cascade $c$ can then be easily represented as a $N$-dimensional vector $\tb^c:=(t_1^c,\ldots,t_N^c)^\top$, where $t_i^c$ is the time when contagion $c$ reaches node $i$, and $t_i^c \in [0, T] \cup \cbr{\infty}$.
We repeatedly split all cascades into a 80\% training set and a 20\% test set at random for five times. 
On each training set, we learn the continuous-time model using \netrate~\cite{DBLP:netrate} with exponential transmission functions. For discrete-time model, we learn the infection probabilities using~\cite{Netrapalli:2012:LGE:2254756.2254783} for IC, \spm and \pmia. Similarly, for LT, we follow the methodology by~\cite{kleinberg_kdd03}. Let $\Ccal(u)$ be the set of all cascades where $u$ was
the source node.
%Each cascade $c\in\Ccal(u)$ is an $N$-dimensional vector $\tb^c:=(t_1^c,\ldots,t_N^c)^\top$ with $i$-th dimension recording the infection time of node $i$.  Given an observation window $T$, if a node $i$ is not observed to be %infected by $T$, then $t_i^c=\infty$. Therefore,
%Given an observation window $T$, for each cascade $c\in\Ccal(u)$, the size of the set of nodes $\cbr{v|t_{v}^c< T}$ reflects the real influence of node $u$ up to time $T$.
%
Based on $\Ccal(u)$, the total number of distinct nodes infected before $T$ quantifies the real influence of node $u$ up to time $T$.
In Figure~\ref{real_influence}(a), we report the Mean Absolute Error (MAE) between the real and the estimated influence. Clearly, \continmax performs the best statistically. Because the length of real cascades empirically conforms to a power-law distribution where most cascades are very short (2-4 nodes), the gap of the estimation error is relatively not large. However, we emphasize that such accuracy improvement is critical for maximizing long-term influence. The estimation error for individuals will accumulate along the spreading paths. Hence, any consistent improvement in influence estimation can lead to significant improvement to the overall influence estimation and maximization task, which is further confirmed by Figures~\ref{real_influence}(b) and~\ref{real_influence}(c) where we evaluate the influence of the selected nodes in the same spirit as influence estimation: the true influence is calculated as the total number of distinct nodes infected before $T$ based on $\Ccal(u)$ of the selected nodes. The selected sources given by \continmax achieve the best performance as we vary the number of selected sources and the observation time window.

%\manuel{Here you talk about models and methods exchangeably, and all the description is pretty
%confusing. I rewrote parts trying to clarify it, but I think it needs substantially more work}.
%\begin{figure}
%\vspace{-2mm}
% \centering
% \begin{tabular}{cc}
%\includegraphics[width=0.45\columnwidth]{influence_vs_size_quotes_n600_e1206-crop} &
%\includegraphics[width=0.45\columnwidth]{influence_vs_time_quotes_n600_e1206-crop} \\
%(a) Influence vs. size & Influence vs. time\\
%\end{tabular}
%  \vspace{-3mm}
%\caption{\label{real_influence} Influence $\sigma(\Acal, T)$ against : (a) the source size by $T=5$ and (b) the time window with $|\mathcal{A}|=20$.}
% \vspace{-4mm}
%\end{figure}

\vspace{-3.5mm}
\section{Conclusions}
\vspace{-3mm}
We propose a randomized influence estimation algorithm in continuous-time diffusion networks, which can scale up to networks of millions of nodes while significantly
improves over previous state-of-the-arts in terms of the accuracy of the estimated influence and the quality of the selected nodes in maximizing the influence. There are also many venues for future work. It will be
interesting to apply the current algorithm to other tasks like influence minimization and manipulation, and design scalable algorithms for continuous-time models other than the independent cascade model.
\vspace{-1mm}

{{\bf Acknowledgement:} Our work is supported by NSF/NIH BIGDATA 1R01GM108341-01,  NSF IIS1116886, NSF IIS1218749, a DARPA Xdata grant and Raytheon Faculty Fellowship of Gatech.}
\clearpage
\newpage

\bibliographystyle{unsrt}
\bibliography{nan}

\clearpage
\newpage

\begin{appendix}

\section{Heterogeneous Transmission Functions}
\label{app:transmission}

We denote the waiting time distribution, or transmission function, along a directed edge of $\Gcal$ as $f_{ji}(t_i|t_j)$. Formally, the transmission function $f_{ji}(t_i|t_j)$ for directed edge
$j\rightarrow i$ is the conditional density of node $i$ getting infected at time $t_i$ given that node $j$ was infected at time $t_j$. We assume it is shift invariant, \ie, $f_{ji}(t_i|t_j) =
f_{ji}(t_i - t_j) = f_{ji}(\tau_{ji})$, where $\tau_{ji}:=t_i - t_j$, and it takes positive values when $\tau_{ji} \geq 0$, and the value of zero otherwise.

In most previous work, simple parametric transmission functions such as the exponential distribution $\alpha_{ji}\exp(-\alpha_{ji} \tau_{ji})$, and the Rayleigh distribution
$\alpha_{ji} \tau \exp(-\alpha_{ji} \tau_{ji}^2 / 2)$ have been used~\cite{survival,DBLP:netrate}.
However, in many real world scenarios, information transmission between pairs of nodes can be heterogeneous and the waiting times can obey distributions that dramatically differ from these
simple models. For instance, in viral marketing, active consumers could update their status instantly, while an inactive user may just log in and respond once a day. As a result, the transmission
function between an active user and his friends can be quite different from that between an inactive user and his friends.
As an attempt to model these complex scenarios, nonparametric transmission functions have been recently considered~\cite{nan_nips2012}. In such approach, the relationship between the
survival function, the conditional intensity function or hazard, and the transmission function is exploited.
In particular, the survival function is defined as $S_{ji}(\tau_{ji}) := 1 - \int_{0}^{\tau_{ji}}f_{ji}(\tau')d\tau'$ and the hazard function is defined as $h_{ji}(\tau_{ji}) := f_{ji}(\tau_{ji})/S_{ji}(\tau_{ji})$.
Then, it is a well-known result in survival theory that $S_{ji}(\tau_{ji}) = \exp\rbr{-\int_0^{\tau_{ji}} h_{ji}(\tau') d\tau'}$ and $f_{ji}(\tau_{ji}) = h_{ji}(\tau_{ji}) S_{ji}(\tau_{ji})$.
The advantage of using the conditional intensity function is that we do not need to explicitly enforce ``the integral equals 1'' constraint for the conditional density $f_{ji}$. Instead, we just need to
ensure $h_{ji} \geq 0$. This facilitates nonparametric modeling of the transmission function. For instance, we can define the conditional intensity function as a positive combination of $n$ positive
kernel functions $k$,
\begin{align*}
 h_{ji}(\tau) = \sum\nolimits_{l = 1}^n \alpha_l k(\tau_l,\tau),~\text{if $\tau > 0$, and $0$ otherwise}.
\end{align*}
A common choice of the kernel function is the Gaussian RBF kernel $k(\tau',\tau) = \exp(-\nbr{\tau-\tau'}^2/2s^2)$. Nonparametric transmission functions significantly improve
modeling of real world diffusion, as is shown in \cite{nan_nips2012}.

\section{A Graphical Model Perspective}
\label{app:graphicalmodel}

Now, we look at the independent cascade model from the perspective of graphical models, where the collection of random variables includes the infection times $t_i$ of the nodes. Although the
original contact graph $\Gcal$ can contain directed loops, each diffusion process (or a cascade) induces a directed acyclic graph (DAG). For those cascades consistent with a particular
DAG, we can model the joint density of $t_i$ using a directed graphical model:
\begin{align}
	p\rbr{\{t_i\}_{i\in\Vcal}} = \prod\nolimits_{i \in \Vcal} p\rbr{t_i | \{t_j\}_{j \in \pi_i}},
	\label{eq:dag_factorization}
\end{align}
where each $\pi_i$ denotes the collection of parents of node $i$ in the induced DAG, and each term $p(t_i | \{t_j\}_{j \in \pi_i})$ corresponds to a conditional density of $t_j$ given the infection times
of the parents of node $i$. This is true because given the infection times of node $i$'{}s parents, $t_i$ is independent of other infection times, satisfying the local Markov property of a directed graphical
model. We note that the independent cascade model only specifies explicitly the pairwise transmission functions for each directed edge, but does not directly define the conditional density
$p(t_i | \{t_j\}_{j \in \pi_i})$.

However, these conditional densities can be derived from the pairwise transmission functions based on the Independent-Infection property~\cite{DBLP:netrate}:
\begin{align}
  \hspace{-2mm}
	p\rbr{t_i | \{t_j\}_{j \in \pi_i}} = \sum\nolimits_{j\in\pi_i}h_{ji}(t_i|t_j)\prod\nolimits_{l\in\pi_{i}}S(t_i|t_l),
	\label{eq:cpt_transmission}
\end{align}
which is the sum of the likelihoods that node $i$ is infected by each parent node $j$. More precisely, each term in the summation can be interpreted as the instantaneous risk of
node $i$ being infected at $t_i$ by node $j$ given that it has survived the infection of all parent nodes until time $t_i$.

Perhaps surprisingly, the factorization in Eq.~\eq{eq:dag_factorization} is the same factorization that can be used for an arbitrary induced DAG consistent with the contact
network $\Gcal$. In this case, we only need to replace the definition of $\pi_i$ (the parent of node $i$ in the DAG)  to the set of neighbors of node $i$ with an edge pointing to node $i$
in $\Gcal$. This is not immediately obvious from Eq.~\eq{eq:dag_factorization}, since the contact network $\Gcal$ can contain directed loops which may be in conflict with the conditional
independence semantics of directed graphical models. The reason it is possible to do so is as follows: Any fixed set of infection times, $t_1,\ldots,t_d$, induces an ordering of the infection
times. If $t_i \leq t_j$ for an edge $j\rightarrow i$ in $\Gcal$, $h_{ji}(t_i|t_j)=0$, and the corresponding term in Eq.~\eq{eq:cpt_transmission} is zeroed out, making the conditional
density consistent with the semantics of directed graphical models.

Based on the joint density of the infection times in Eq.~\eq{eq:dag_factorization}, we can perform various inference and learning tasks. For instance, previous work has used Eq.~\eq{eq:dag_factorization}
for learning the parameters of the independent cascade model~\cite{nan_nips2012,DBLP:netrate,infopath}. However, this may not be the
most convenient form for addressing other inference problems, including the influence estimation problem in the next section. To this end, we propose an alternative view.

Instead of directly modeling the infection times $t_i$, we can focus on the collection of mutually independent random transmission times $\tau_{ji} = t_i - t_j$. In this case, the joint density of the collection
of transmission times $\tau_{ji}$ is fully factorized
\begin{align*}
	p\rbr{\{\tau_{ji}\}_{(j,i)\in \Ecal}} = \prod\nolimits_{(j,i)\in\Ecal} f_{ji}(\tau_{ji}),
\end{align*}
where $\Ecal$ denotes the set of edges in the contact network $\Gcal$ --- switching from
the earlier node-centric view to the now edge-centric view. Based on the Shortest-Path property of the independent cascade model, variable $t_i$ can be viewed
as a transformation from the collection of variables $\{\tau_{ji}\}_{(j,i)\in \Ecal}$. More specifically, let $\Qcal_i$ be the collection of directed paths in $\Gcal$ from the source nodes to node
$i$, where each path $q\in \Qcal_i$ contains a sequence of directed edges $(j,l)$, and assuming all source nodes are infected at zero time, then we obtain variable $t_i$ via
\begin{align}
 t_i = g_i\rbr{\{\tau_{ji}\}_{(j,i)\in \Ecal}} := \min_{q \in \Qcal_i} \sum\nolimits_{(j,l)\in q} \tau_{jl},
\end{align}
where $g_i(\cdot)$ is the transformation.
% , and the shortest path $q_i^\ast := \argmin_{q \in \Qcal_i} \sum_{(j,l)\in q} \tau_{jl}$.

Importantly, we can now compute the probability of infection of node $i$ at $t_i$ using the set of variables $\{\tau_{ji}\}_{(j,i)\in\Ecal}$:
\begin{align}
	\Pr\cbr{t_i \leq T} = \Pr\cbr{g_i\rbr{\{\tau_{ji}\}_{(j,i)\in \Ecal}} \leq T}.
 	\label{eq:equivalence1}
\end{align}
%
% This will be very useful when we try to tackle the influence estimation problem discussed in the next section. \manuel{Several times you use, it will be useful in the future... but not now,
% produces a bit of impatient feeling}.
The significance of the relation is that it allows us to transform a problem involving a sequence of dependent variables $\{t_i\}_{i\in\Vcal}$ to one
with independent variables $\{\tau_{ji}\}_{(j,i)\in\Ecal}$. Furthermore, the two problems are connected via the shortest path algorithm in weighted directed graph, a standard well studied
operation in graph analysis.

% Suppose that for all nodes $v_i\in\Vcal$, we have a sequence of infection time stamps $\tb = (t_0,...,t_{n-1})$. If some node $k$ survives during the diffusion process, we define $t_k = \infty$. We assume that after node $i$ is infected by one of its parents, it will be not infected again by other parents. Because any parent $j\in\pi_i$ could be the first one to infect $i$, the likelihood of observing such a sequence is
% \begin{align}
% \ell\rbr{\tb} & = \prod_i\rbr{\sum_{j\in\pi_i}f_{ji}(t_i|t_j)\prod_{k\in\pi_{i},k\neq j}S(t_i|t_j)} \nonumber\\
% & = \prod_i\rbr{\sum_{j\in\pi_i}h_{ji}(t_i|t_j)\prod_{k\in\pi_{i}}S(t_i|t_j)}
% \end{align}
% Let $p(t_i | \cbr{t_j}_{j\in\pi_i}) = \sum_{j\in\pi_i}h_{ji}(t_i|t_j)\prod_{k\in\pi_{i}}S(t_i|t_j)$. We can treat $p(t_i | \cbr{t_j}_{j\in\pi_i})$ as the conditional probabilities of $t_i$ given the time stamps of all its parents, in other words, the likelihood factorizes similarly to a directed graphical model when $\Gcal$ is also a DAG as
% \begin{align}
% \ell\rbr{\tb} = \prod_{i}p(t_i | \cbr{t_j}_{j\in\pi_i})
% \end{align}
% Then, the probability that node $i$ is infected within a time window $[0,T]$ given some node $1$ as the source is simply derived as
% \begin{align}
% &P\cbr{t_i\in[0,T]} = \int_{t_2 = 0}^\infty\dotsm\int_{t_3 = 0}^\infty\dotsm\int_{t_i=0}^T\dotsm\int_{t_{n-1}=0}^\infty\nonumber\\
% & \ell(0,t_{2},\dotsc,t_{n-1})dt_2dt_3\dotsc dt_i\dotsc dt_{n-1}
% \label{infection_probability}
% \end{align}

\section{Naive Sampling Algorithm}
\label{app:ns}

The graphical model perspective described in Section~\ref{sec:graphicalmodel} and Appendix~\ref{app:graphicalmodel} suggests a naive sampling (NS) algorithm for approximating $\sigma(\Acal,T)$:
\begin{tight_list}
  \item[1.] Draw $n$ samples, $\cbr{\cbr{\tau_{ji}^l}_{(j,i)\in\Ecal}}_{l=1}^n$,~\iid~from the waiting time product distribution $\prod_{(j,i) \in \Ecal} f_{ji}(\tau_{ji})$;
  \item[2.] For each sample $\cbr{\tau_{ji}^l}_{(j,i)\in\Ecal}$ and for each node $i$, find the shortest path from source nodes to node $i$; count the number of nodes with $g_i\rbr{\cbr{\tau_{ji}^l}_{(j,i)\in\Ecal}}\leq T$;
  \item[3.] Average the counts across $n$ samples.
\end{tight_list}

Although the naive sampling algorithm can handle arbitrary transmission function, it is not scalable to networks with millions of nodes. We need to compute the shortest path for each node and each sample, which
results in a computational complexity of $O(n |\Ecal| + n |\Vcal|\log|\Vcal|)$ for a single source node. The problem is even more pressing in the influence maximization problem, where we need to estimate the influence
of source nodes at different location and with increasing number of source nodes. To do this, the algorithm needs to be repeated, adding a multiplicative factor of $C |\Vcal|$ to the computational complexity ($C$ is
the number of nodes to select). Then, the algorithm becomes quadratic in the network size. When the network size is in the order of thousands and millions, typical in modern social network analysis, the naive sampling
algorithm become prohibitively expensive. Additionally, we may need to draw thousands of samples ($n$ is large), further making the algorithm impractical for large scale problems.

\section{Least Label List}
\label{app:leastlabellist}

The notation ``$\text{argsort}((r_1,\ldots,r_{|\Vcal|}),\text{ascend})$'' in line 2 of Algorithm~\ref{a1} means that we sort the collection of random labels in ascending order and return the argument of the sort as an
ordered list.

\begin{algorithm}[h]
	\SetAlgoVlined
	\KwIn{a reversed directed graph $\Gcal=(\Vcal, \Ecal)$ with edge weights $\{\tau_{ji}\}_{(j,i)\in\Ecal}$, a node labeling $\cbr{r_i}_{i\in \Vcal}$}
	\KwOut{A list $r_\ast(s)$ for each $s\in \Vcal$}

	\lFor{each $s\in\Vcal$}{$d_s\leftarrow\infty, r_\ast(s)\leftarrow\emptyset$}

	\For{$i$ in $\text{argsort}((r_1,\ldots,r_{|\Vcal|}),\text{ascend})$}{
		empty heap ${\tt H}\leftarrow \emptyset$\;

		set all nodes except $i$ as unvisited\;

		push $(0,i)$ into heap ${\tt H}$\;

		\While{${\tt H} \ne\emptyset$}{
			pop $(d_\ast, s)$ with the minimum $d_\ast$ from ${\tt H}$\;

			add $(d_\ast, r_i)$ to the end of list $r_\ast(s)$\;

			$d_s\leftarrow d^\ast$\;

			\For{each unvisited in-coming neighbor $j$ of $s$}{
                set $j$ as visited\;

				\uIf{$(d, j)$ in heap ${\tt H}$}{
					Pop $(d, j)$ from heap ${\tt H}$\;

					Push $(\min\cbr{d,d_\ast + \tau_{js}}, j)$ into heap ${\tt H}$\;
				}
				\ElseIf{$d_\ast + \tau_{js} < d_j$}{
					Push $(d_\ast + \tau_{js}, j)$ into heap ${\tt H}$\;
				}
			}
		}
	}
	\caption{Least Label List}
	\label{a1}
\end{algorithm}

\begin{figure}
\begin{minipage}[b]{0.45\textwidth}
\includegraphics[width=1\textwidth,height=95pt]{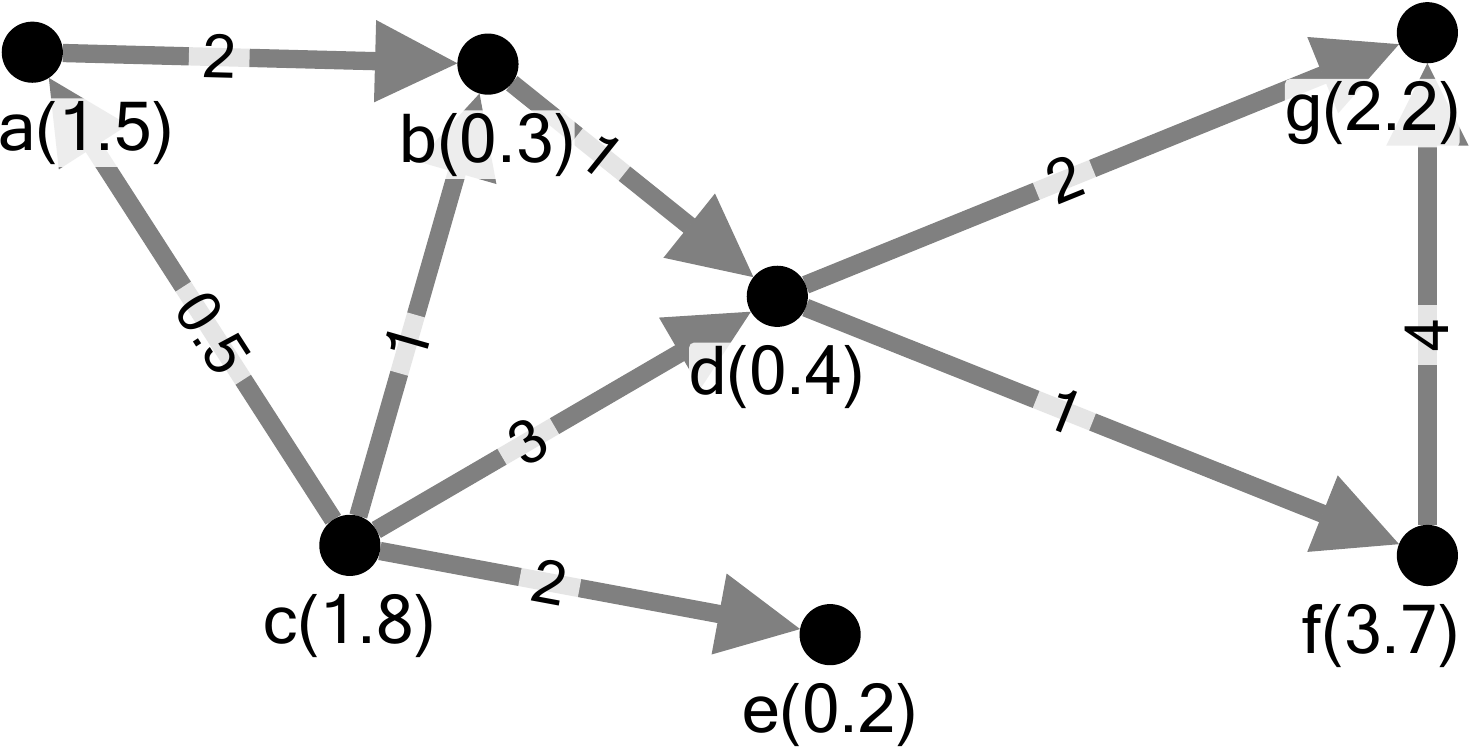}
\end{minipage}
\hspace{0.5cm}
\begin{minipage}[b]{0.45\textwidth}
{\small
  \begin{tabular}{l}
  $\bullet$ Node labeling : \\
  ~~~~$e(0.2) < b(0.3) < d(0.4) < a(1.5) < c(1.8) < g(2.2) < f(3.7)$\\\\
  $\bullet$ Neighborhoods: \\
%   $\Ncal(c,0.5)=\{a,c\};~\Ncal(c,1)=\{a,b,c\};$\\
  ~~~~$\Ncal(c,2)=\{a,b,c,e\};~\Ncal(c,3)=\{a,b,c,d,e,f\};$\\\\
%   $\Ncal(c,4)=\{a,b,c,d,e,f,g\};$\\
  $\bullet$ Least-label list: \\
  ~~~~$r_\ast(c): (2,0.2),(1,0.3),(0.5,1.5),(0,1.8)$ \\\\
  $\bullet$ Query: $r_\ast(c,0.8) = r(a) = 1.5$ \\

  \end{tabular}
  }
\end{minipage}
\caption{\label{demo}Graph $\Gcal=(\Vcal,\Ecal)$, edge weights $\cbr{\tau_{ji}}_{(j,i)\in\Ecal}$, and node labeling $\cbr{r_i}_{i\in\Vcal}$ with the associated output from Algorithm~\ref{a1}.}
 \end{figure}

Figure~\ref{demo} shows an example of the Least-Label-List. The nodes from $a$ to $g$ are assigned to exponentially distributed labels with mean one shown in each parentheses. Given a query distance 0.8 for node $c$, we can binary-search its Least-label-list $r_*(c)$ to find that node $a$ belongs to this range with the smallest label $r(a) = 1.5$.

\newpage

\section{Theorem 1}
\label{app:proof1}
\newtheorem{thm2}{Theorem}

\begin{thm2}
	Sample the following number of sets of random transmission times
	\begin{align*}
		n \geqslant \frac{C \Lambda}{\epsilon^2} \log\rbr{\frac{2 |\Vcal|}{\delta}}
	\end{align*}
	where $\Lambda:= \max_{\Acal:\abr{\Acal}\leq C} 2\sigma(\Acal,T) / (m-2) + 2Var(S_\tau) + 2 a \epsilon /3$, and for each set of random transmission times, sample $m$ set of random labels,
	we can guarantee that
	$$
		\abr{\widehat{\sigma}(\Acal,T) - \sigma(\Acal,T)} \leqslant \epsilon
	$$
	simultaneously for all $\Acal$ with $\abr{\Acal}\leqslant C$, with probability at least $1 - \delta$.
\end{thm2}
\begin{proof}
	Let $S_\tau:=\abr{\Ncal(\Acal,T)}$ for a fixed set of $\{\tau_{ji}\}$ and then $\sigma(\Acal,T) = \EE_{\tau}[S_\tau]$. The randomized algorithm with $m$ randomizations produces an unbiased estimator $\Shat_\tau=(m-1)/(\sum_{u=1}^m r_\ast^u)$ for $S_\tau$,~\ie,~$\EE_{r|\tau}[\Shat_\tau]=S_\tau$, with variance $\EE_{r|\tau}[(\Shat_\tau - S_\tau)^2] = S_\tau / (m-2)$.

	Then $\Shat_\tau$ is also an unbiased estimator for $\sigma(\Acal,T)$, since $\EE_{\tau,r}[\Shat_\tau] = \EE_{\tau} \EE_{r|\tau} [\Shat_\tau] = \EE_{\tau} [S_\tau] = \sigma(\Acal,T)$. Its variance is
	\begin{align*}
		Var(\Shat_\tau)
		&:=\EE_{\tau,r}[(\Shat_\tau - \sigma(\Acal,T))^2] = \EE_{\tau,r}[(\Shat_\tau - S_\tau + S_\tau - \sigma(\Acal,T))^2] \\
		&= \EE_{\tau,r}[(\Shat_\tau - S_\tau)^2] + 2\, \EE_{\tau,r}[(\Shat_\tau - S_\tau)(S_\tau - \sigma(\Acal,T))] + \EE_{\tau,r}[(S_\tau - \sigma(\Acal,T))^2] \\
		&= \EE_{\tau}[S_\tau / (m-2)] + 0\, + Var(S_\tau) \\
		&= \sigma(\Acal,T) / (m-2) + Var(S_\tau)
	\end{align*}

	Then using Bernstein's inequality, we have, for our final estimator $\widehat{\sigma}(\Acal,T) = \frac{1}{n}\sum_{l=1}^{n} \Shat_{\tau^l}$, that
	\begin{align}
		\Pr\cbr{\abr{\widehat{\sigma}(\Acal,T) - \sigma(\Acal,T)} \geqslant \epsilon } \leqslant  2 \exp\rbr{-\frac{n \epsilon^2}{2 Var(\Shat_\tau) + 2 a \epsilon / 3}}
		\label{eq:boundeddifference}
	\end{align}
	where $\Shat_\tau < a \leqslant |\Vcal|$.

	Setting the right hand side of relation~\eq{eq:boundeddifference}~to $\delta$, we have that, with probability $1 - \delta$, sampling the following number set of random transmission times
	\begin{align*}
		n \geqslant \frac{2 Var(\Shat_\tau) + 2 a \epsilon / 3}{\epsilon^2} \log\rbr{\frac{2}{\delta}}
		= \frac{2\sigma(\Acal,T) / (m-2) + 2Var(S_\tau) + 2 a \epsilon /3}{\epsilon^2} \log\rbr{\frac{2}{\delta}}
	\end{align*}
	we can guarantee that our estimator to have error $\abr{\widehat{\sigma}(\Acal,T) - \sigma(\Acal,T)} \leqslant \epsilon$.

% 	that $\EE[\widehat{\sigma}(\Acal,T)] = \sigma(\Acal,T)$ and
% 	\begin{align*}
% 		Var(\widehat{\sigma})
% 		&:= \EE[(\widehat{\sigma}(\Acal,T) - \sigma(\Acal,T))^2] = \frac{1}{n}Var(\Shat_\tau) \\
% 		&= \sigma(\Acal,T) / (n(m-2)) + Var(S_\tau)/n
% 	\end{align*}

	If we want to insure that $\abr{\widehat{\sigma}(\Acal,T) - \sigma(\Acal,T)} \leqslant \epsilon$ simultaneously hold for all $\Acal$ such that $\abr{\Acal}\leqslant C \ll |\Vcal|$, we can first use union bound with relation~\eq{eq:boundeddifference}. In this case, we have that, with probability $1 - \delta$, sampling the following number set of random transmission times
	\begin{align*}
		n \geqslant \frac{C \Lambda}{\epsilon^2} \log\rbr{\frac{2 |\Vcal|}{\delta}}
	\end{align*}
	we can guarantee that our estimator to have error $\abr{\widehat{\sigma}(\Acal,T) - \sigma(\Acal,T)} \leqslant \epsilon$ for all $\Acal$ with $\abr{\Acal}\leqslant C$.
	Note that we have define the constant $\Lambda:= \max_{\Acal:\abr{\Acal}\leq C} 2\sigma(\Acal,T) / (m-2) + 2Var(S_\tau) + 2 a \epsilon /3$.
\end{proof}

\newpage

\begin{figure}[t]
 \centering
 \renewcommand{\tabcolsep}{0pt}
 \begin{tabular}{ccc}
\includegraphics[width=0.3\textwidth, height = 95pt]{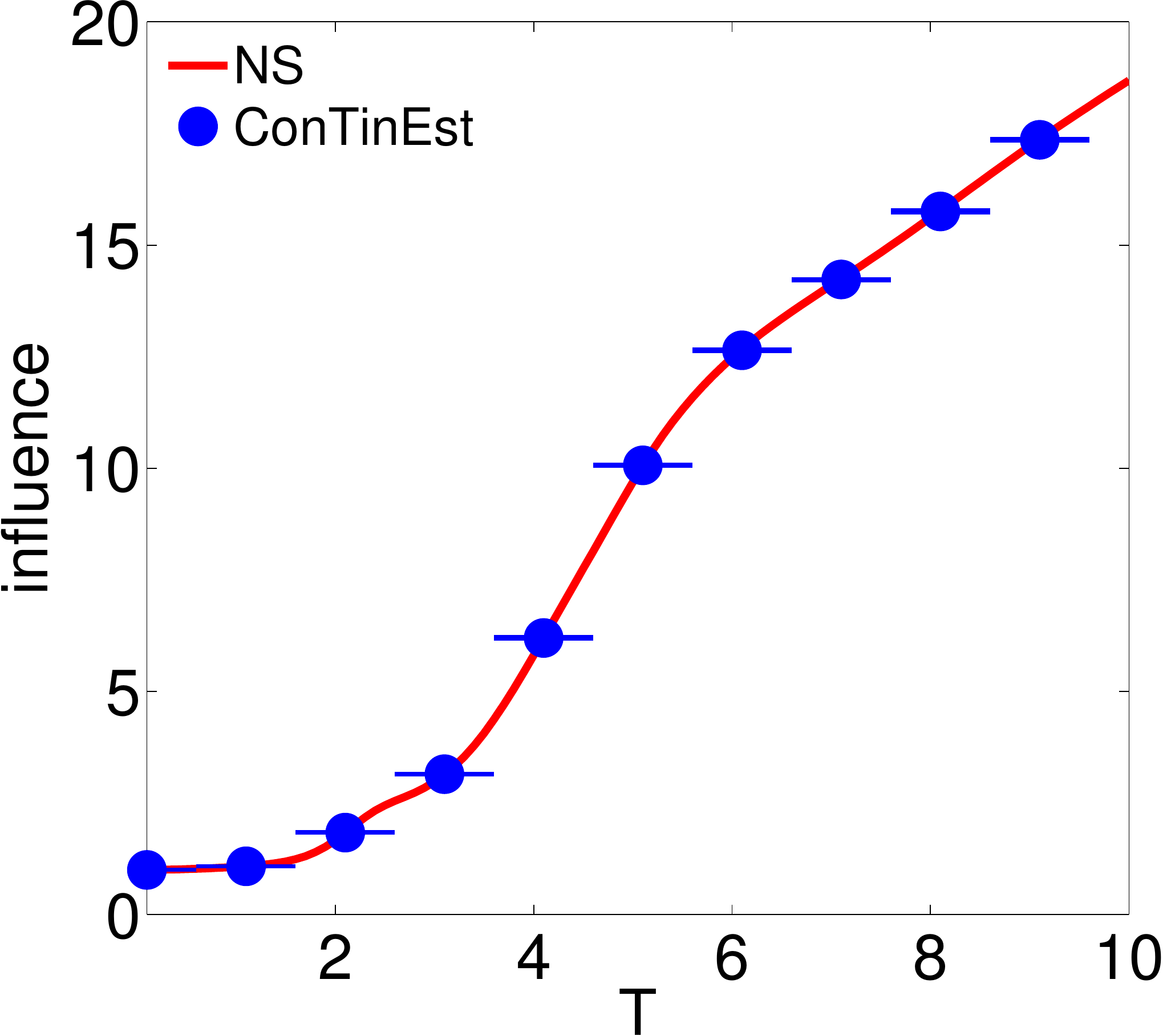} &
\includegraphics[width=0.3\textwidth, height = 95pt]{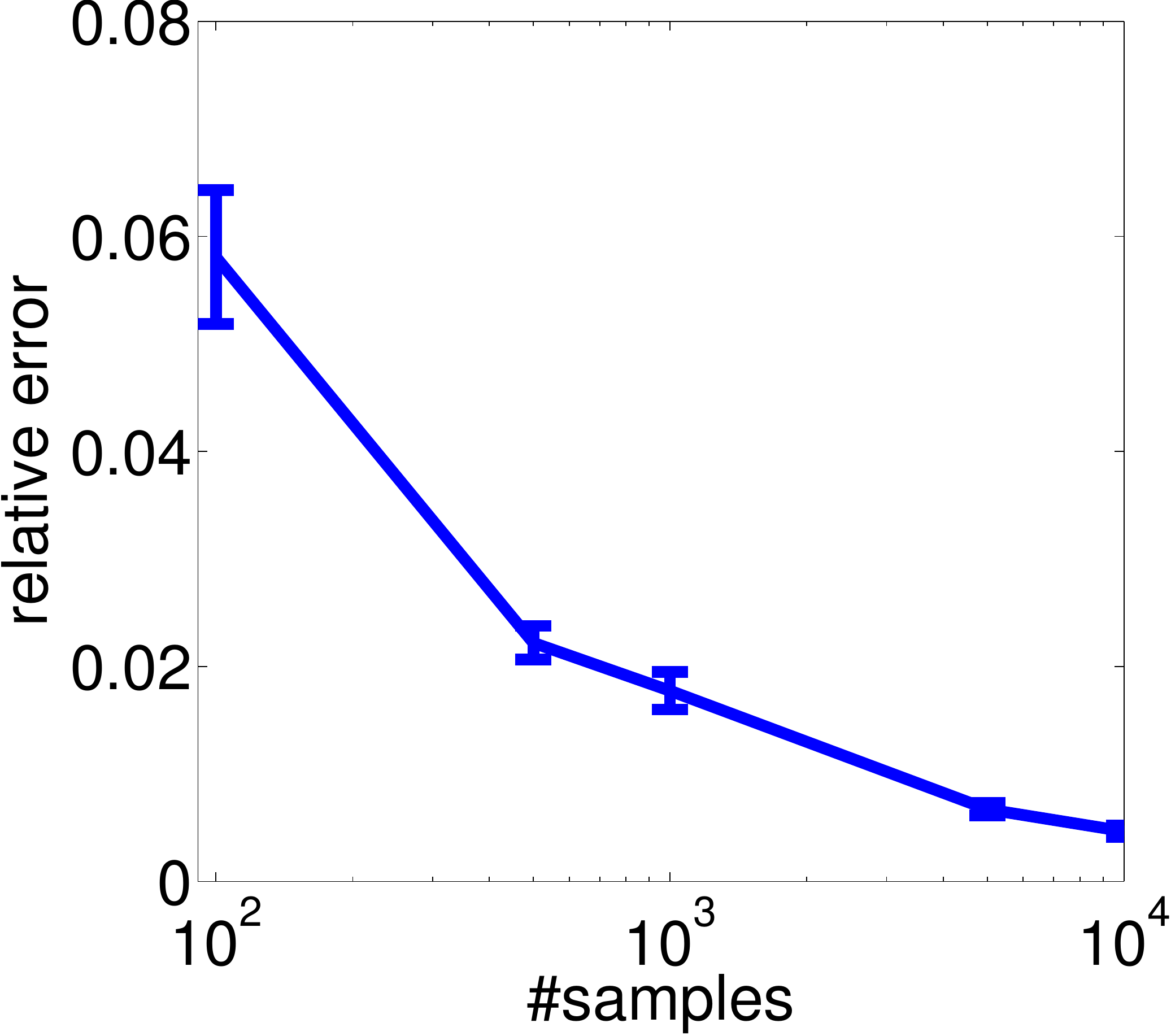} &
\includegraphics[width=0.3\textwidth, height=100pt]{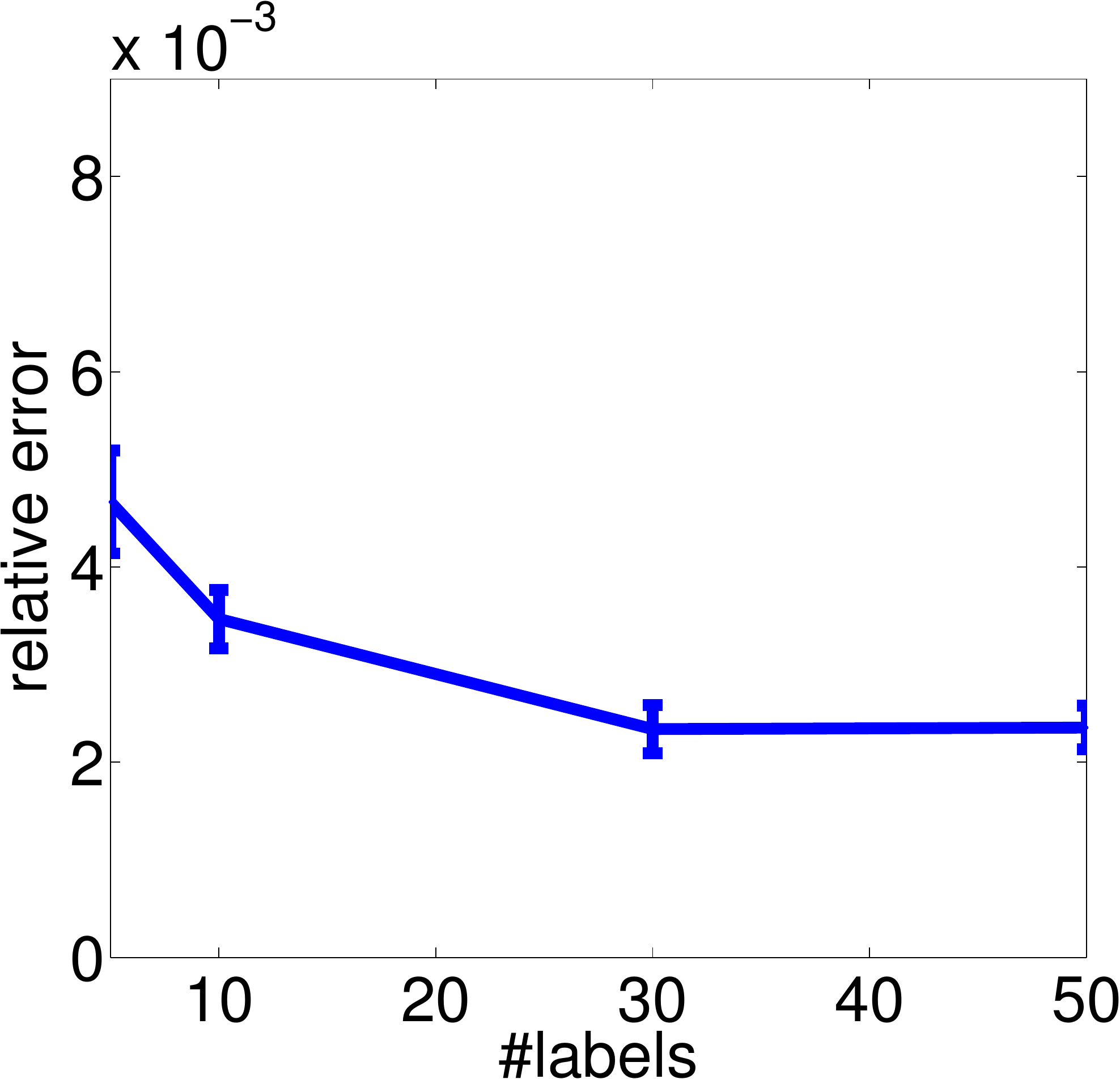} \\
(a) Influence vs. time & (b) Error vs. \#samples & (c) Error vs. \#labels\\
\end{tabular}
 \caption{\label{accuracy_random_wbl} On the {\bf random} kronecker networks with 1,024 nodes and 2,048 edges,
 panels show (a) the estimated influence with increasing time window $T$; (b)  the average relative error for different number of samples, each of which has 5 random labels for every node; and (c) the average relative error for varying number of random labels assigned to every node in each of 10,000 samples. For both (b) and (c), we set  $T=10$. }
\end{figure}
\begin{figure}[h]
 \centering
 \renewcommand{\tabcolsep}{0pt}
 \begin{tabular}{ccc}
\includegraphics[width=0.3\textwidth, height = 95pt]{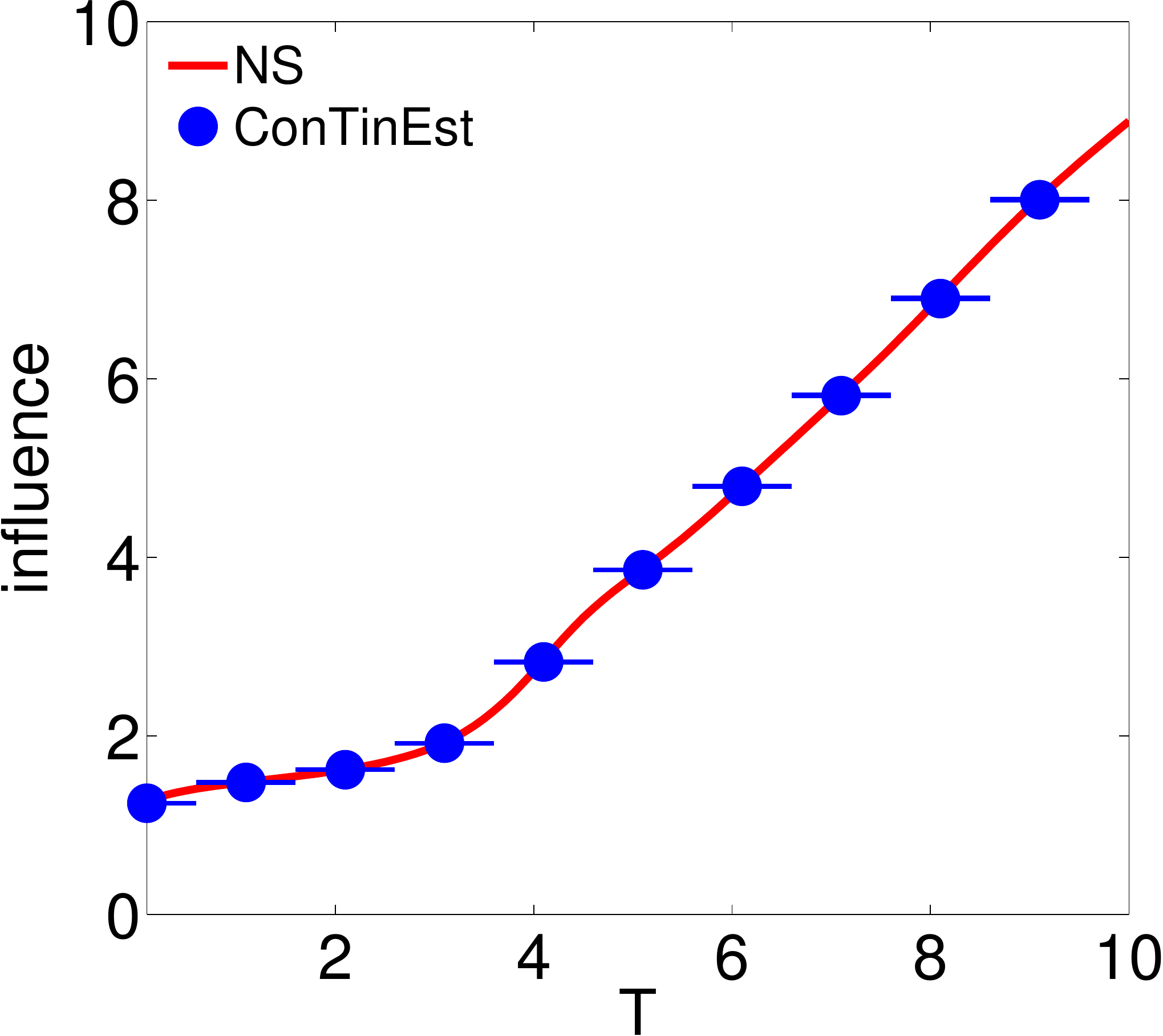} &
\includegraphics[width=0.3\textwidth, height = 95pt]{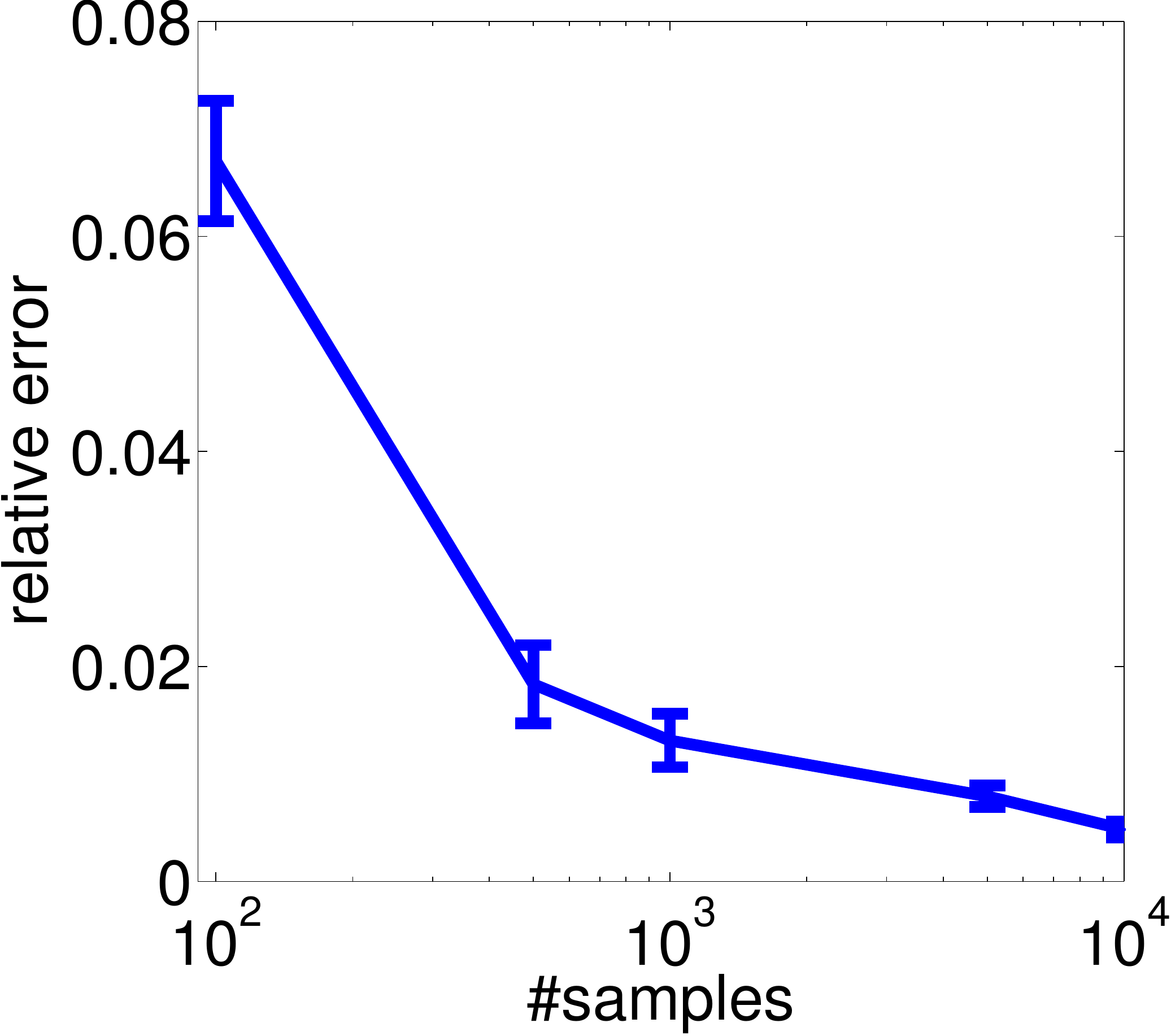} &
\includegraphics[width=0.3\textwidth, height=100pt]{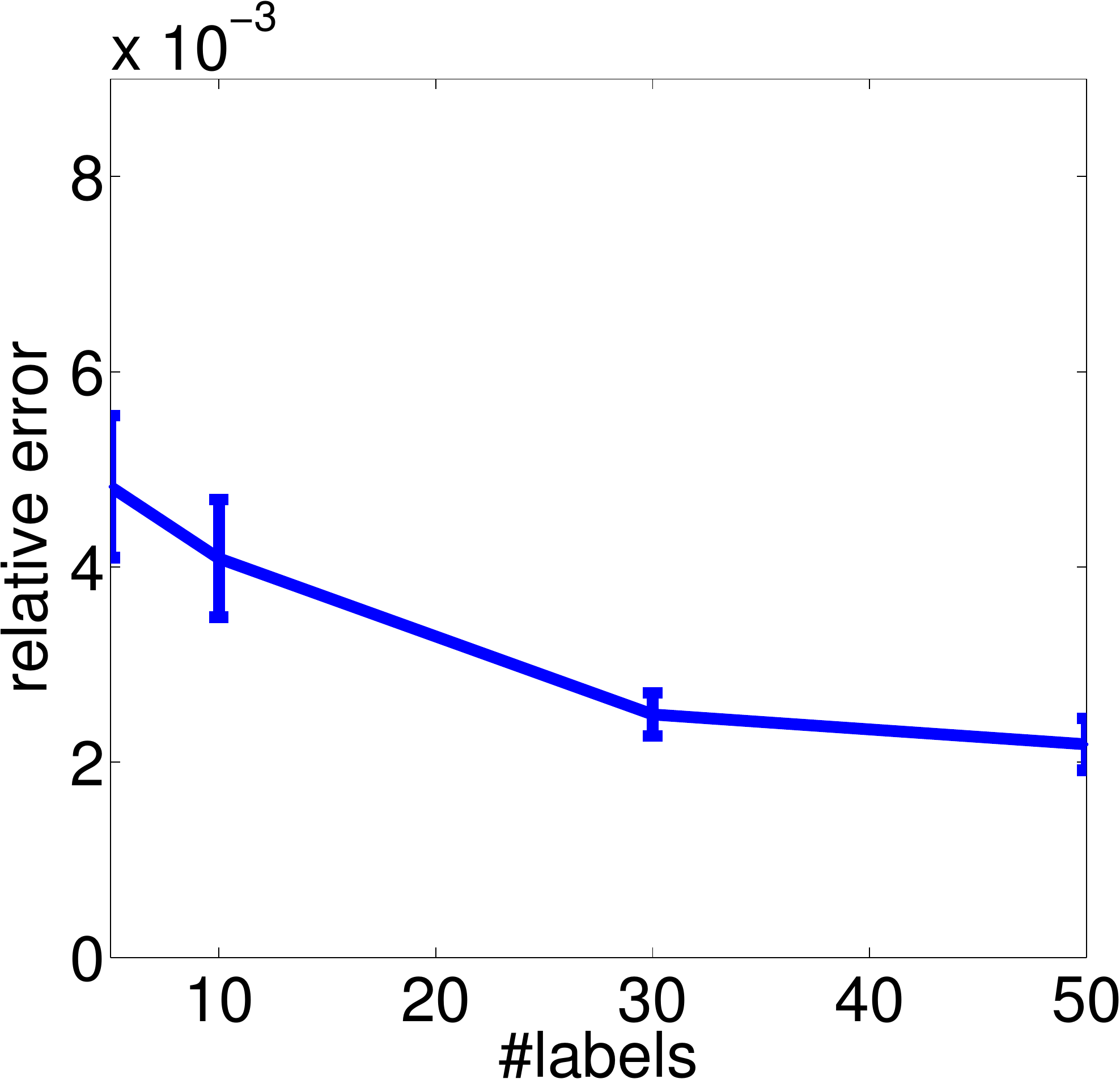} \\
(a) Influence vs. time & (b) Error vs. \#samples & (c) Error vs. \#labels\\
\end{tabular}
 \caption{\label{accuracy_hierarchy_wbl} On the {\bf hierarchical} kronecker networks with 1,024 nodes and 2,048 edges,
 panels show (a) the estimated influence with increasing time window $T$; (b)  the average relative error for different number of samples, each of which has 5 random labels for every node; and (c) the average relative error for varying number of random labels assigned to every node in each of 10,000 samples. For both (b) and (c), we set  $T=10$. }
\end{figure}

\section{Additional Experimental Results}
\label{app:exp}
In this section, we report additional experimental results on accuracy of influence estimation, continuous-time influence maximization and scalability for the synthetic networks.
\subsection{Accuracy of Influence Estimation}
\label{app:accuracy}
Figure~\ref{accuracy_random_wbl} evaluates the estimated scope of influence for different time windows and the relative errors with respective to different number of random samples and labels on the random kronecker
networks with 1,024 nodes and 2,048 edges. Figure~\ref{accuracy_hierarchy_wbl} further reports similar results on the hierarchical kronecker networks. In all cases, the errors decrease dramatically as we draw
more samples and labels.

\begin{figure}[t]
 \centering
 \renewcommand{\tabcolsep}{0pt}
 \begin{tabular}{ccc}
\includegraphics[width=0.3\textwidth, height=95pt]{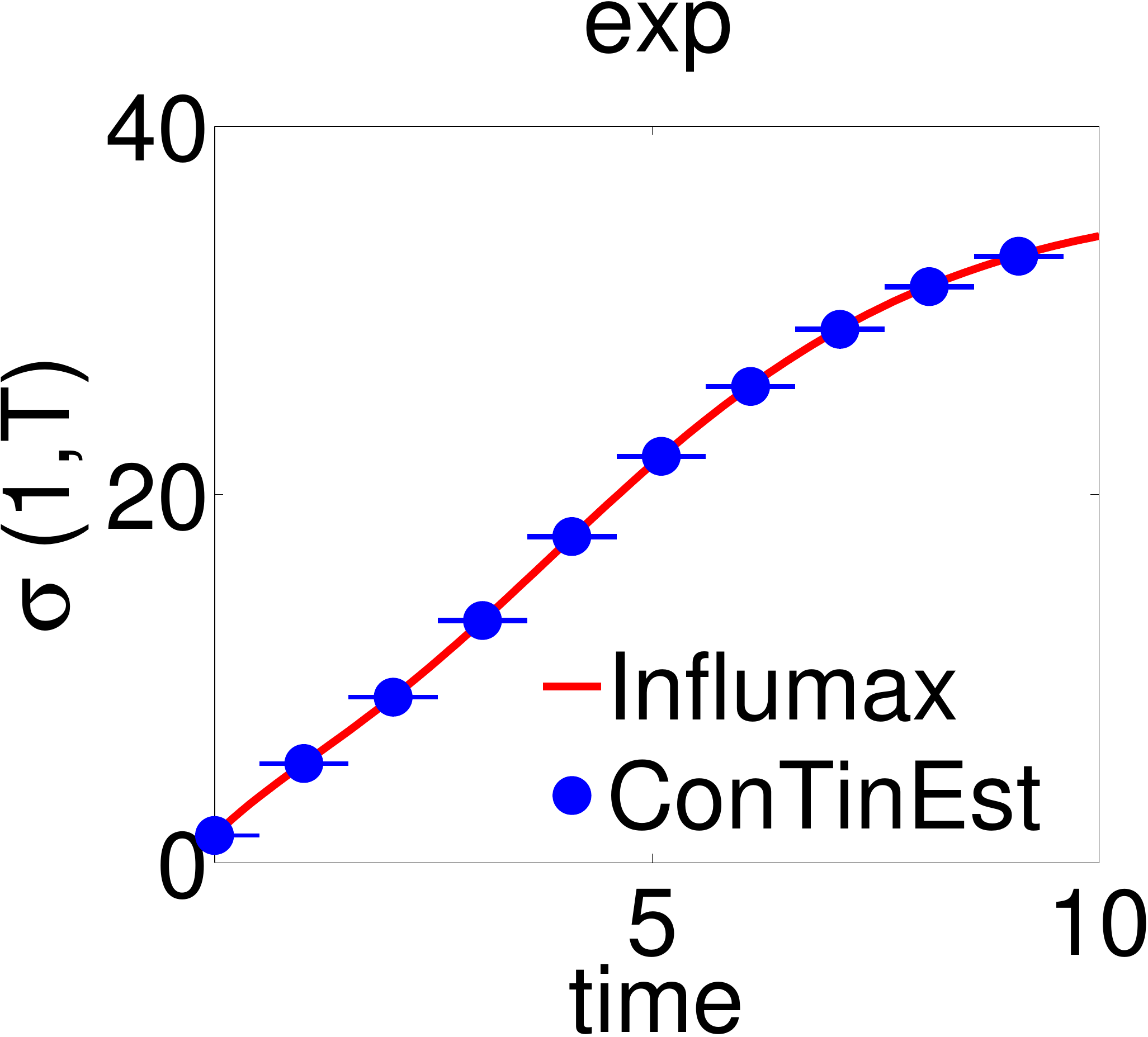} &
\includegraphics[width=0.3\textwidth, height=95pt]{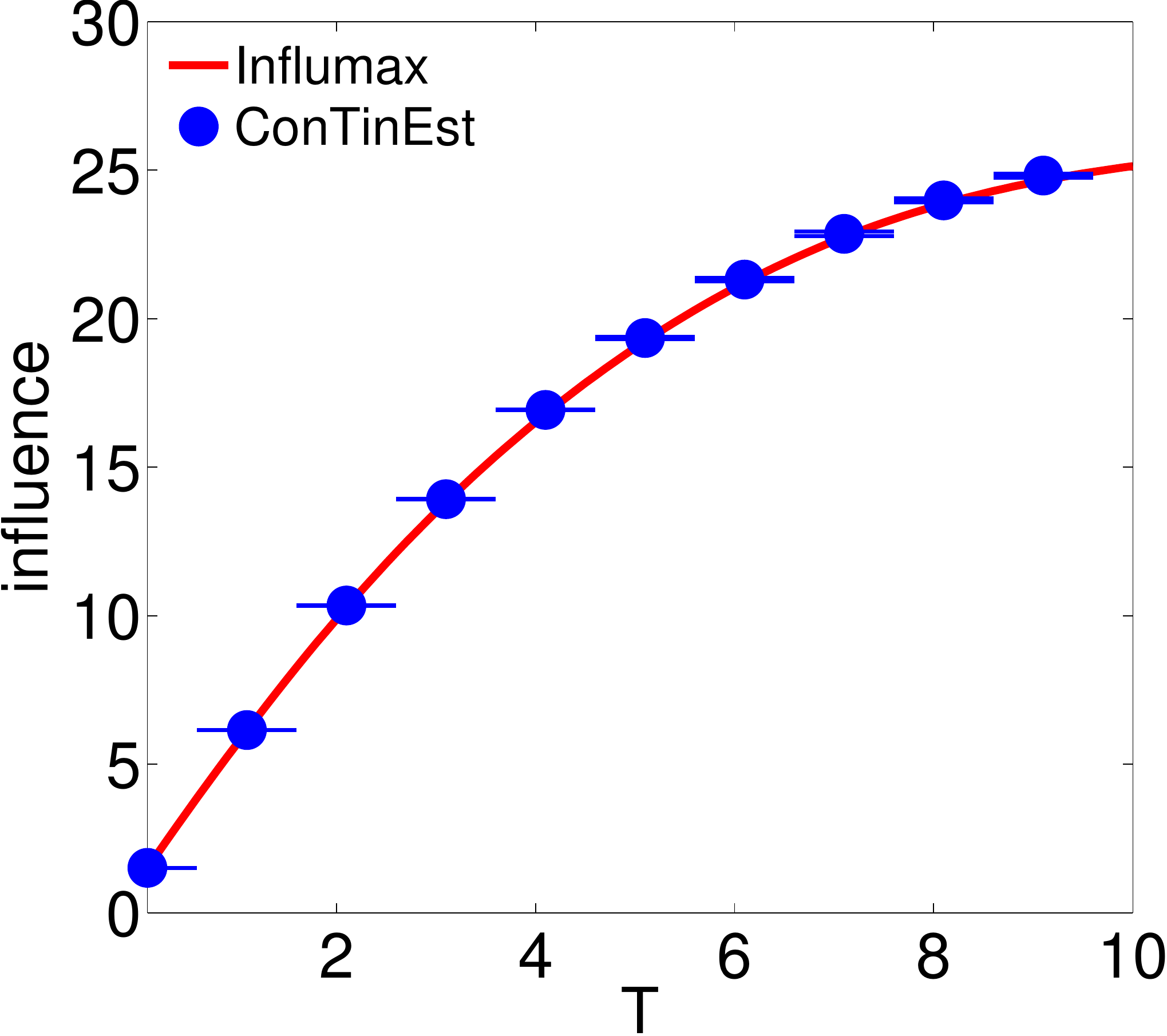} &
\includegraphics[width=0.3\textwidth, height=95pt]{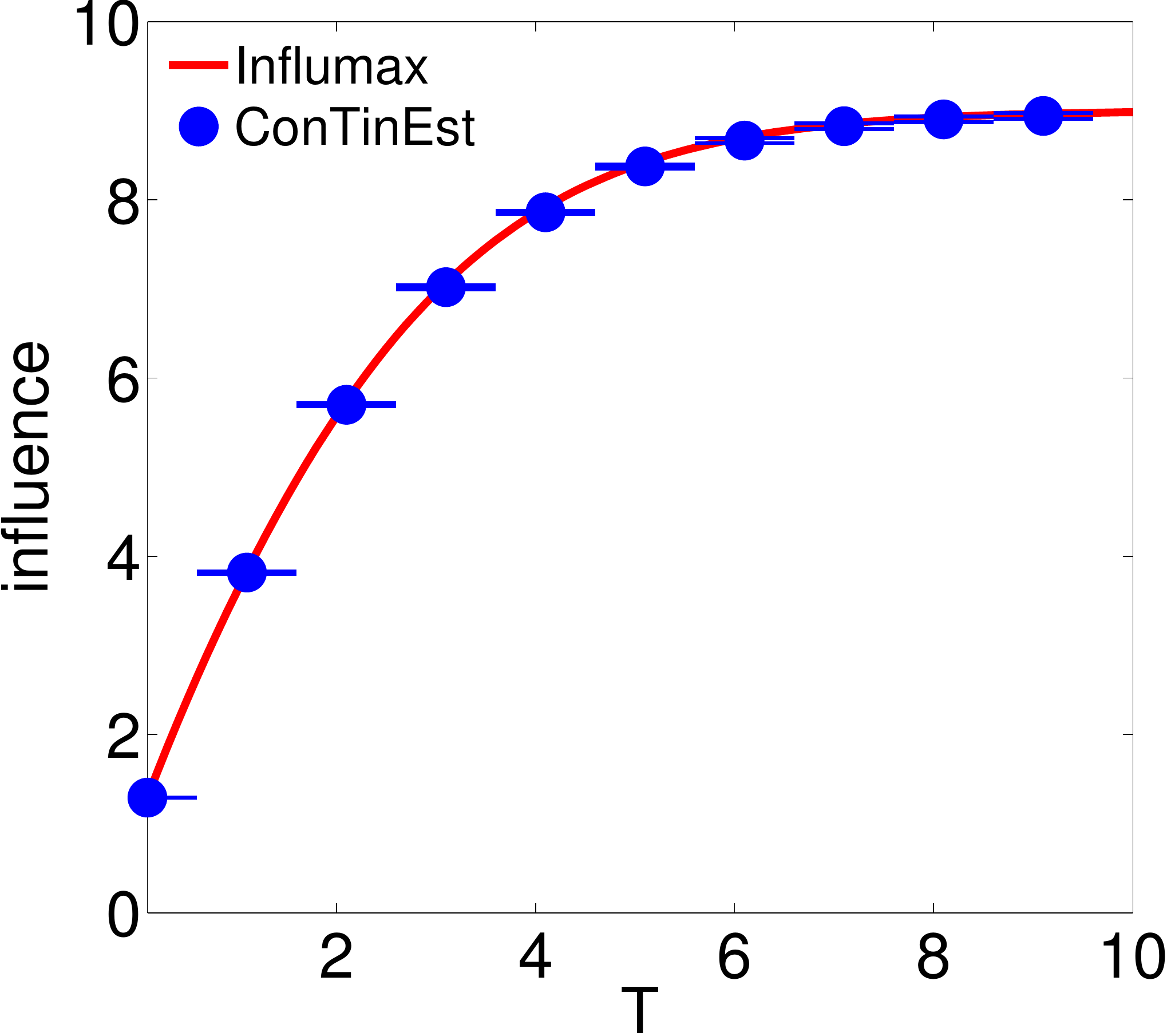} \\
(a) Core-periphery & (b) Random & (c) Hierarchical\\
%\includegraphics[width=0.35\textwidth]{accuracy_outter_core_exp-crop} &
%\includegraphics[width=0.35\textwidth]{accuracy_outter_random_exp-crop} &
%\includegraphics[width=0.35\textwidth]{accuracy_outter_hierarchy_exp-crop} \\
%(d) core & (e) random & (f) hierarchy\\
%\includegraphics[width=0.35\textwidth]{accuracy_inner_core_exp-crop} &
%\includegraphics[width=0.35\textwidth]{accuracy_inner_random_exp-crop} &
%\includegraphics[width=0.35\textwidth]{accuracy_inner_hierarchy_exp-crop} \\
%(g) core & (h) random & (i) hierarchy\\
\end{tabular}
 \caption{\label{accuracy_exp} Infected neighborhood size over three different types of networks with the exponential transmission function associated with each edge. Each type of network consists of 128 nodes and 141 edges. For panels (d-i), we set the observation window $T=10$.}
\end{figure}

In addition, because \influmax can produce exact closed form influence on sparse small networks with exponential transmission functions, we compare \continmax with \influmax in Figure~\ref{accuracy_exp}, where we
chose the highest degree node in the network as the source. We have drawn 10,000 random samples, each of which has 5 random labels for each node. \continmax outputs values of influence which are very close
to the exact values given by \influmax, with relative error less than 0.01 in all three types of networks.

\newpage
 \begin{figure}[t]
	\centering
	 \renewcommand{\tabcolsep}{0pt}
	\begin{tabular}{ccc}
		\includegraphics[width=0.3\textwidth, height = 95pt]{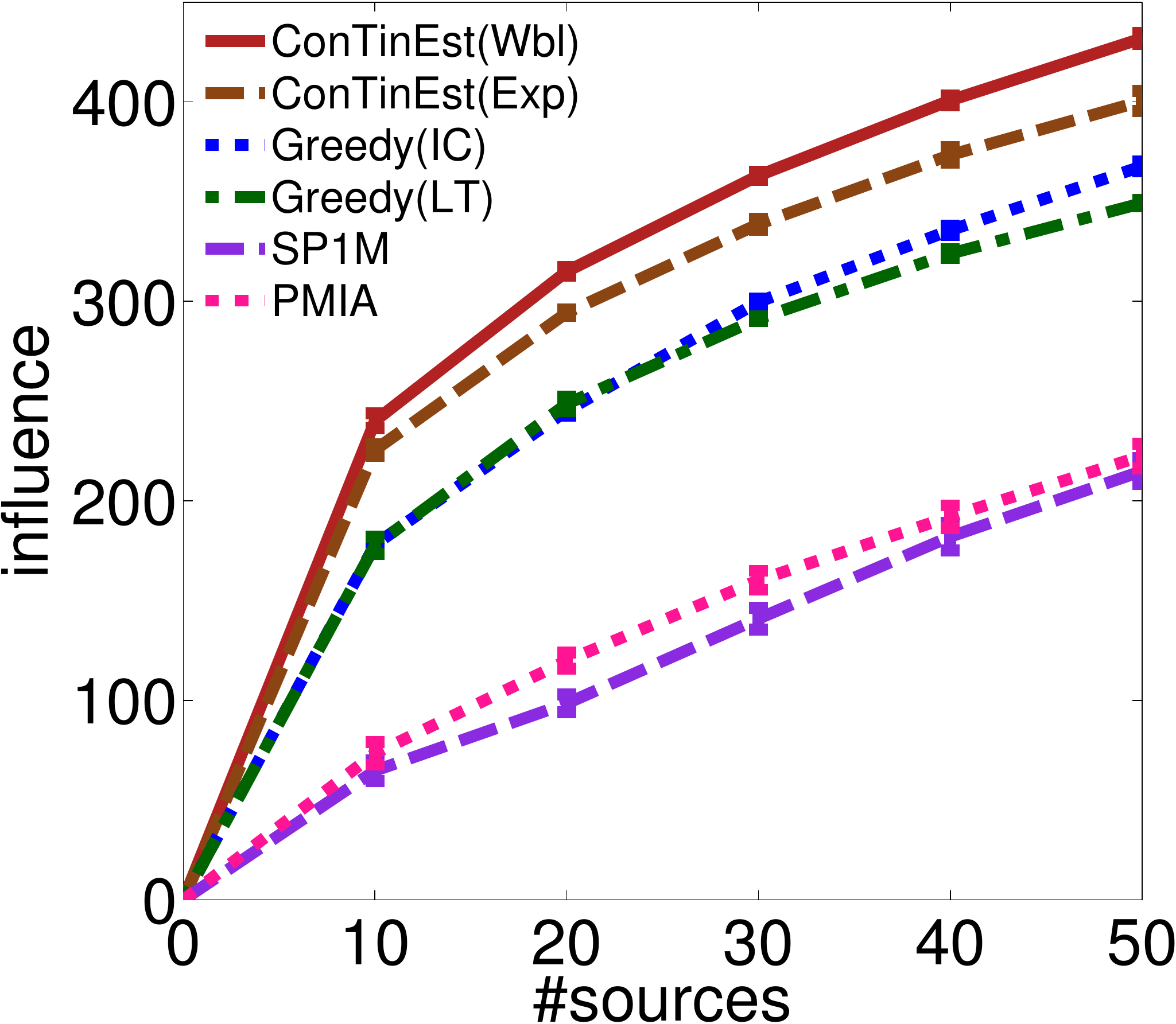} &
		\includegraphics[width=0.3\columnwidth, height = 95pt]{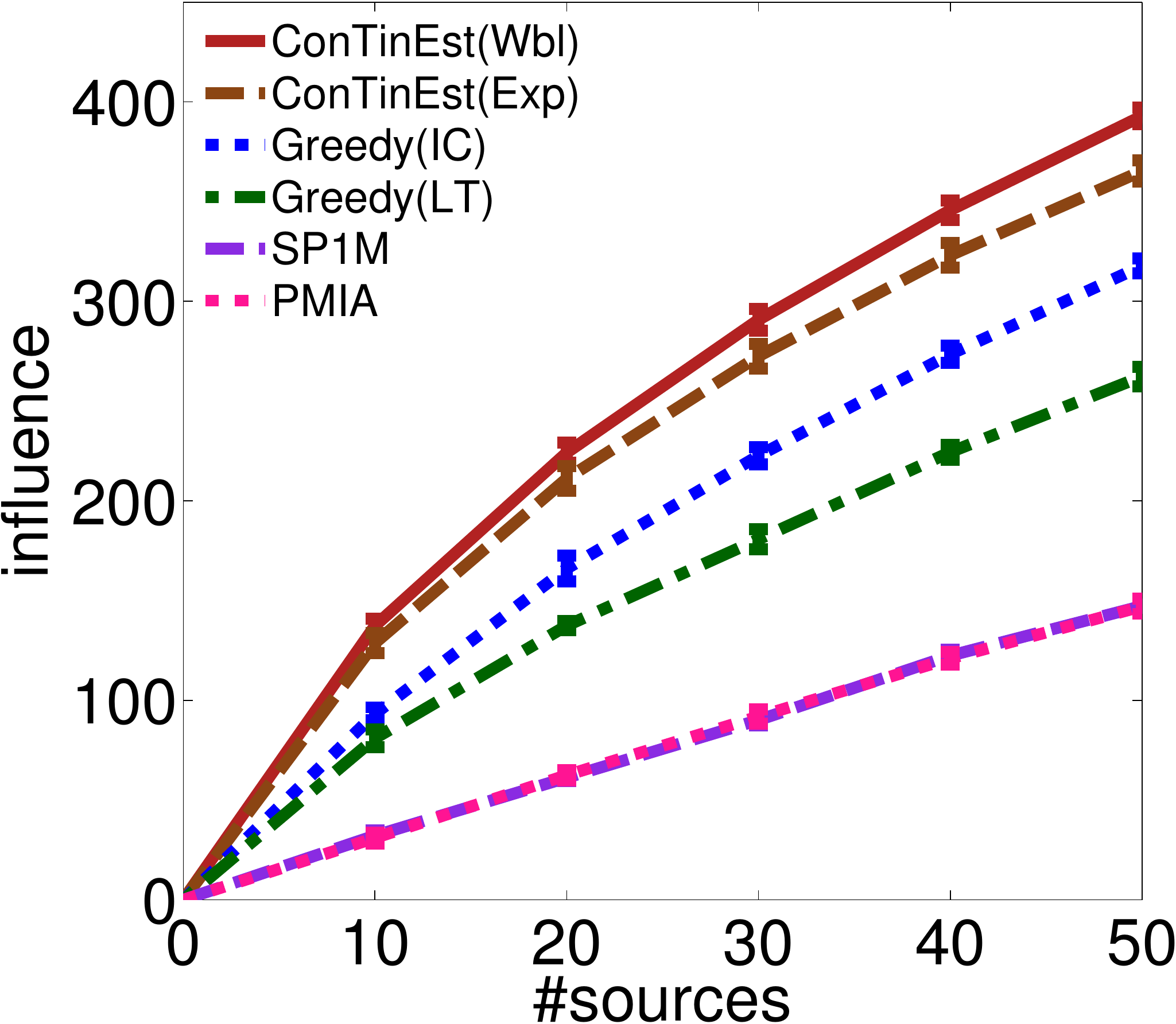}  &
		\includegraphics[width=0.3\columnwidth, height = 95pt]{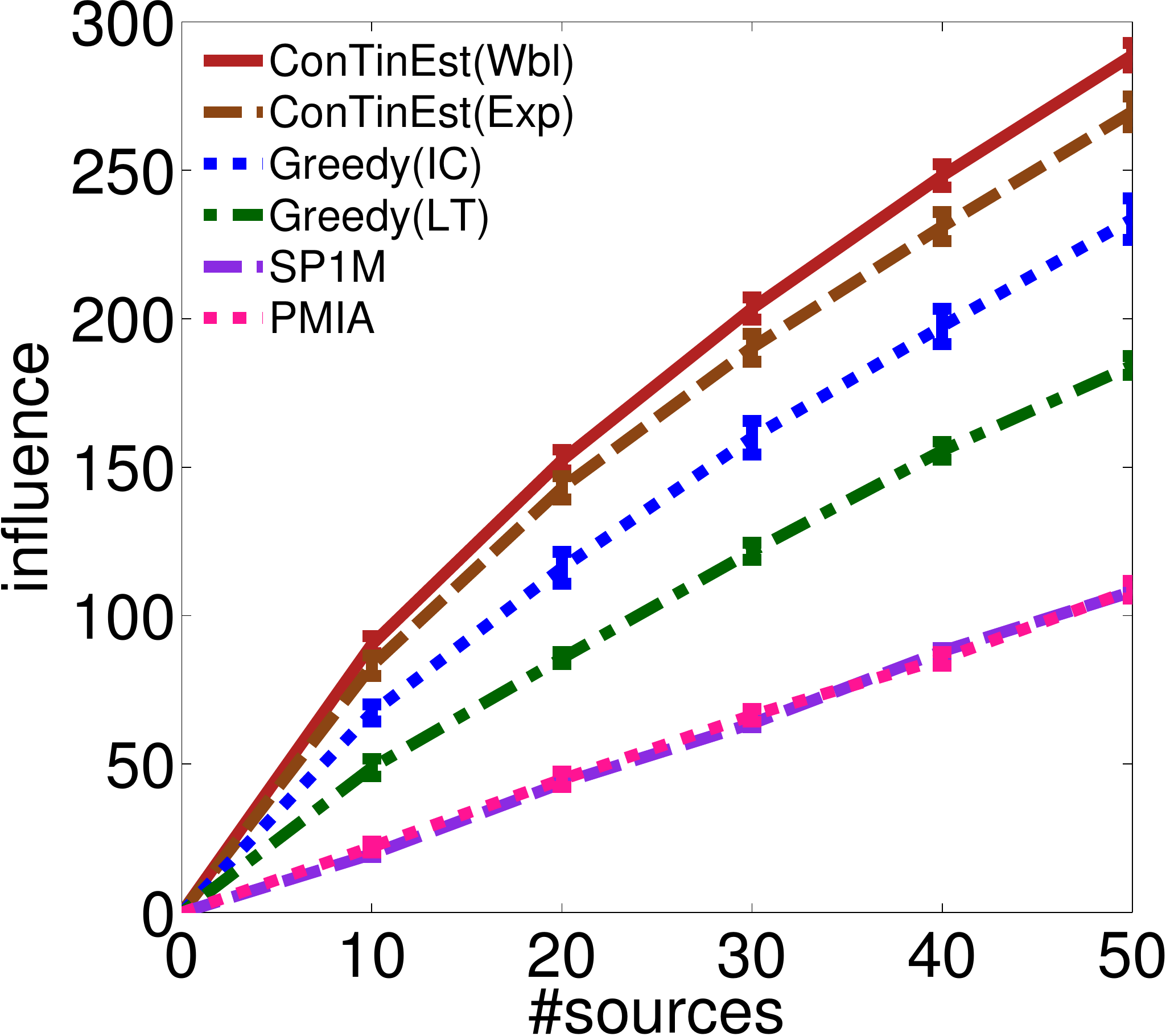}
		\\
		(a) Core-periphery & (b) Random & (c) Hierarchical
	\end{tabular}
	\caption{ \label{influence} Panels present the influence against  the number of sources by $T=5$  on the networks having 1,024 nodes and 2,048 edges with heterogeneous Weibull transmission functions. }
\end{figure}
\begin{figure}[th]
	\centering
	 \renewcommand{\tabcolsep}{0pt}
	\begin{tabular}{ccc}
		\includegraphics[width=0.3\textwidth, height = 95pt]{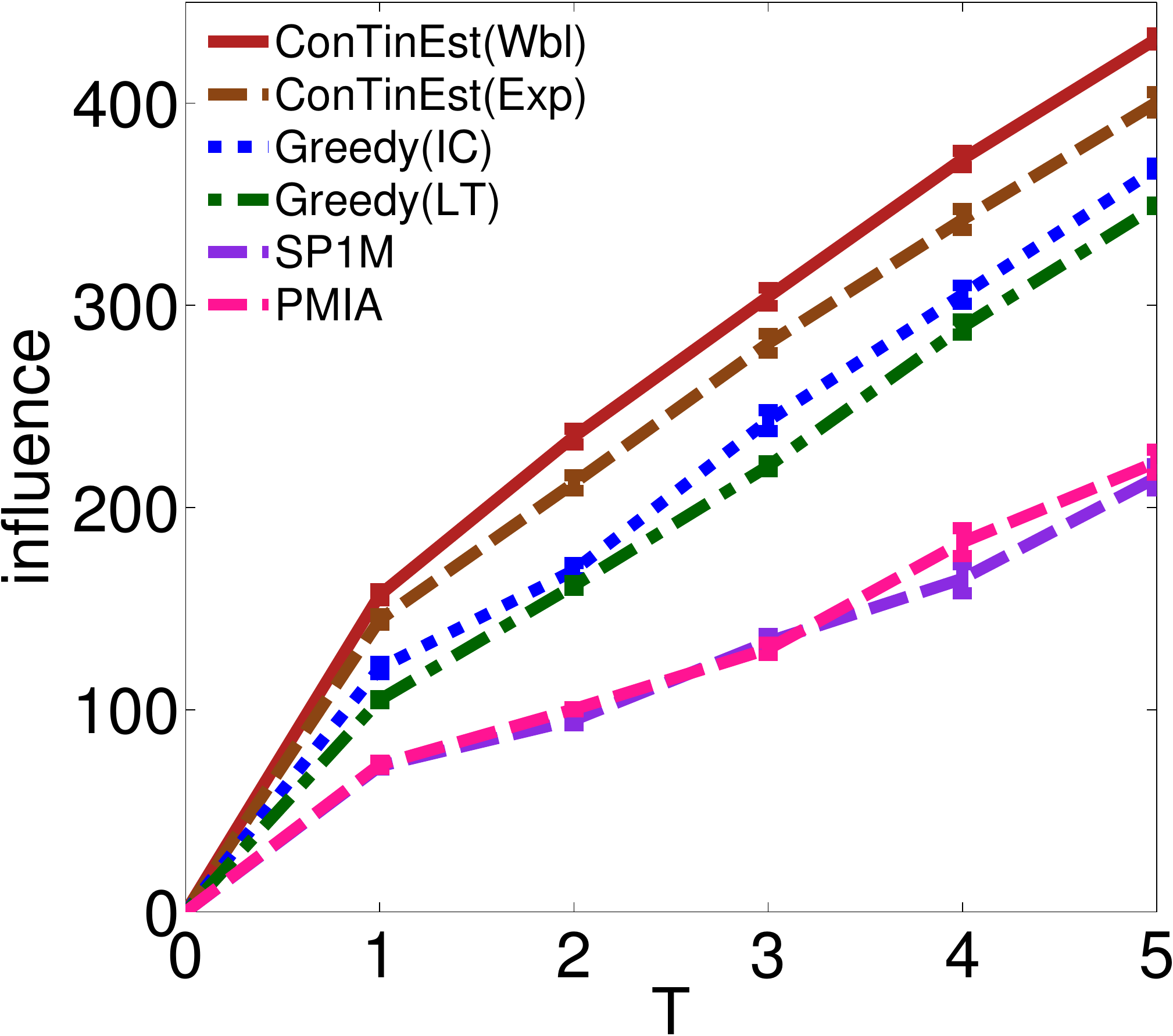} &
		\includegraphics[width=0.3\columnwidth, height = 95pt]{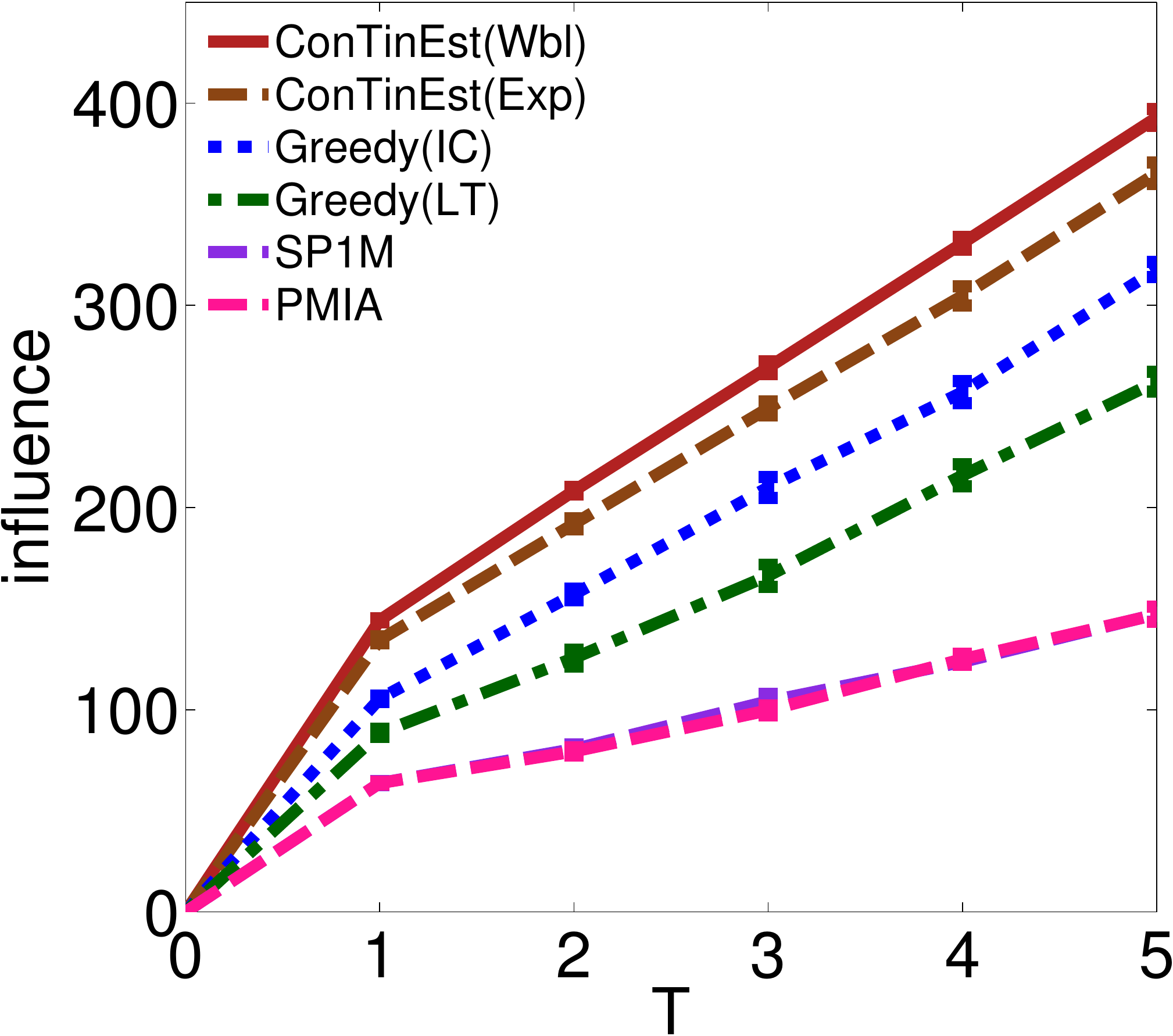}  &
		\includegraphics[width=0.3\columnwidth, height = 95pt]{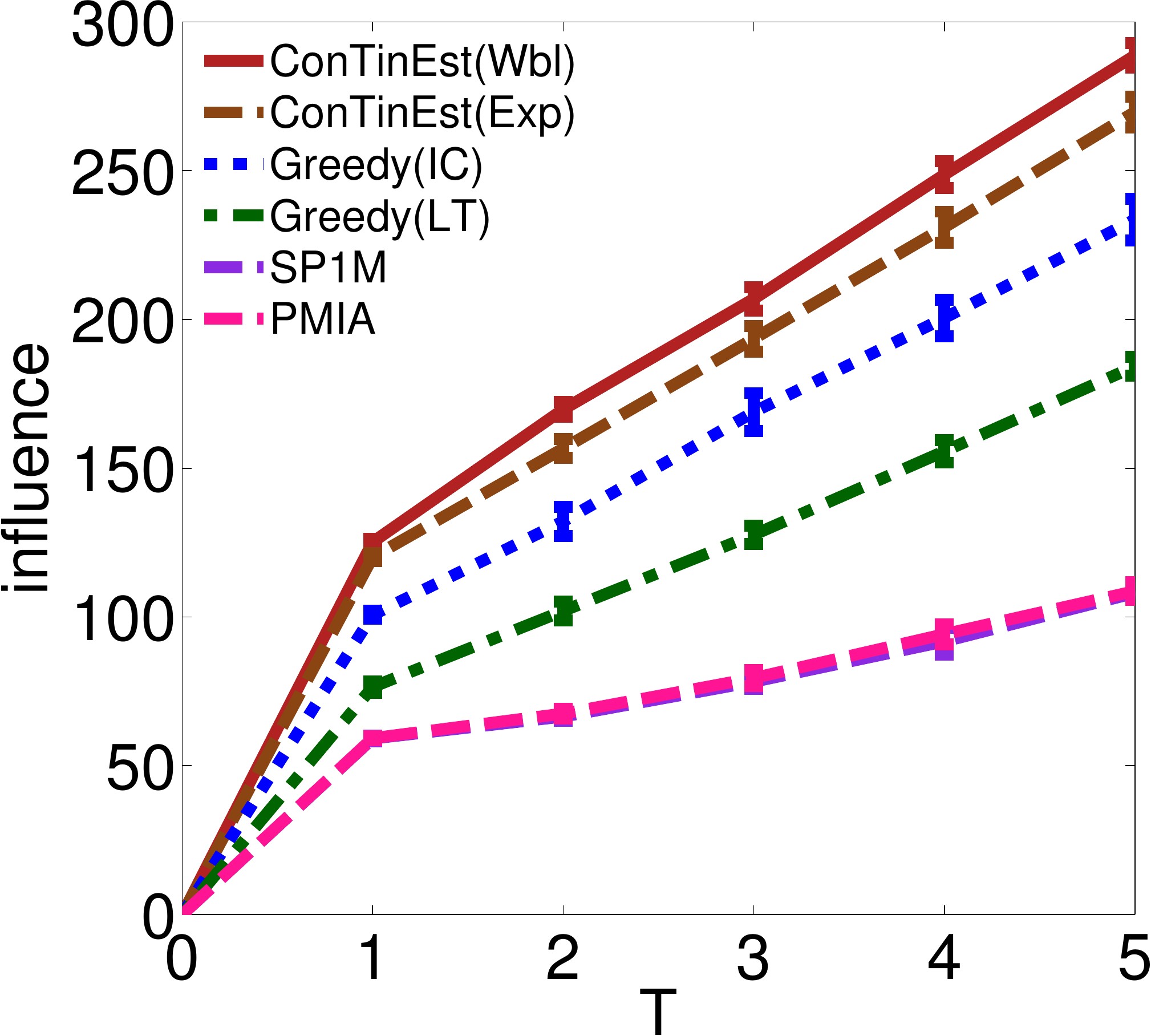}
		\\
		(a) Core-periphery & (b) Random & (c) Hierarchal
	\end{tabular}
	\caption{ \label{influence_time} Panels present the influence against the time window $T$ using 50 sources on the networks having 1,024 nodes and 2,048 edges with heterogeneous Weibull transmission functions. }
\end{figure}

 \subsection{ Continuous-time Influence Maximization}
 \label{app:maximization}
 We compare \continmax to other influence maximization methods based on discrete-time diffusion models: traditional greedy~\cite{kleinberg_kdd03}, with discrete-time Linear Threshold Model (LT) and Independent Cascade Model (IC) diffusion models, and the heuristic methods SP1M~\cite{ChenWY09} and PMIA~\cite{Chen:2010:SIM:1835804.1835934}.
%%We use the core-periphery networks (1,024 nodes, 2,048 edges) used in the previous section, with Weibull pairwise transmission functions.
%%
For \influmax,  since it only supports exponential pairwise transmission functions, we fit an exponential distribution per edge. Furthermore, \influmax is not scalable; when the average network density of the synthetic networks is $\sim2.0$, the run time for \influmax is more than $24$ hours. Instead, we present the results of \continmax using fitted exponential distributions (Exp).
For the discrete-time IC model, we learn the infection probability within time window $T$ using Netrapalli'{}s method~\cite{Netrapalli:2012:LGE:2254756.2254783}. The learned pairwise infection probabilities are also served for \spm and \pmia, which essentially approximately calculate the influence based on the IC model.
For the discrete-time LT model, we set the weight of each incoming edge to a node $u$ to the inverse of its in-degree, as in previous work~\cite{kleinberg_kdd03}, and choose each node's threshold uniformly at random.
Figure~\ref{influence} compares the expected number of infected nodes against source set size for different methods. \continmax outperforms the rest, and the competitive advantage becomes more dramatic the larger the source set grows.
Figure~\ref{influence_time} shows the expected number of infected nodes against the time window for 50 selected sources. Again, \continmax~performs the best for all three types of networks.

\begin{figure}[t]
 \centering
 \renewcommand{\tabcolsep}{0pt}
   \begin{tabular}{cccc}
\includegraphics[width=0.25\textwidth, height = 95pt]{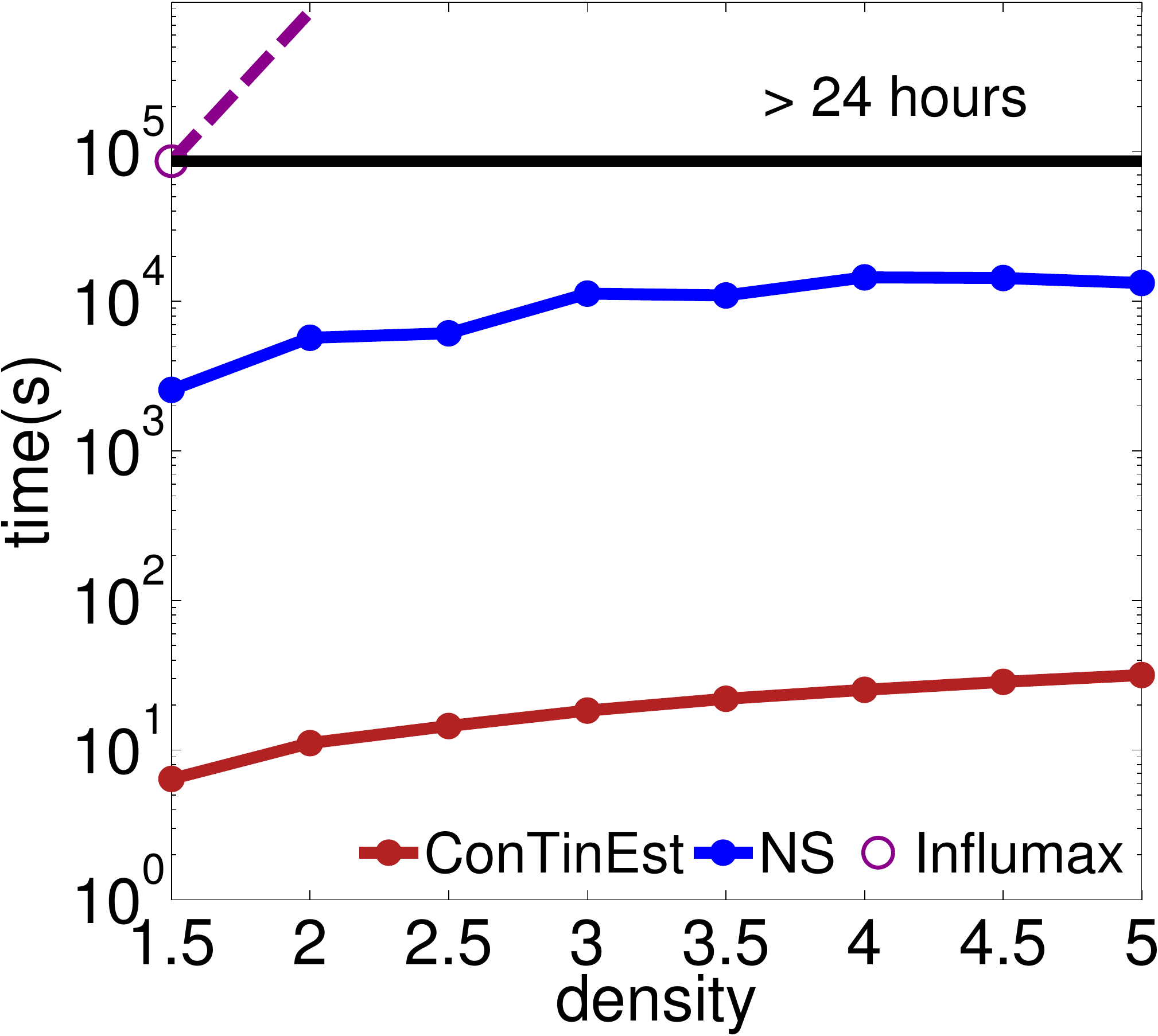} &
\includegraphics[width=0.25\textwidth, height = 95pt]{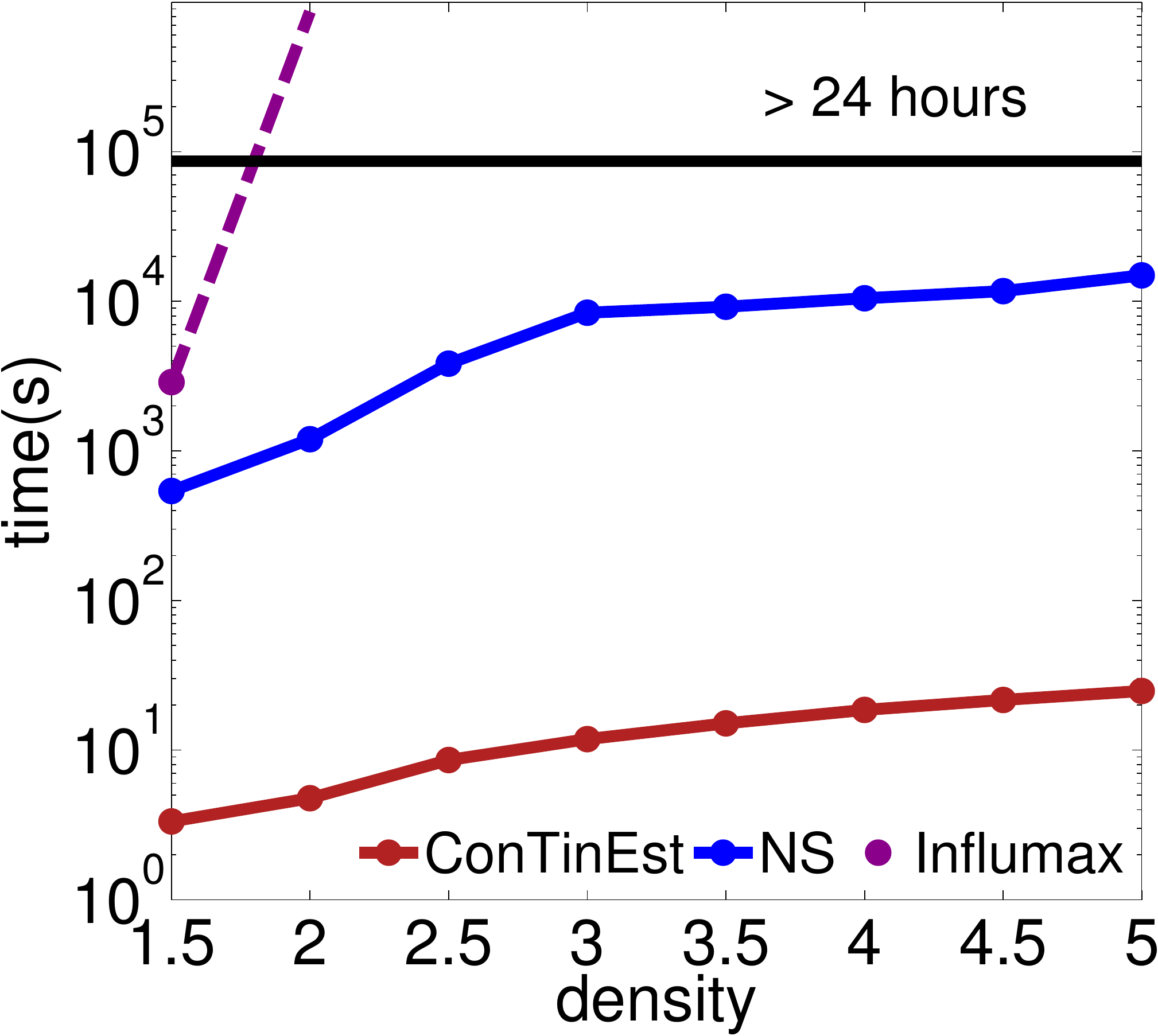} &
\includegraphics[width=0.25\textwidth, height = 95pt]{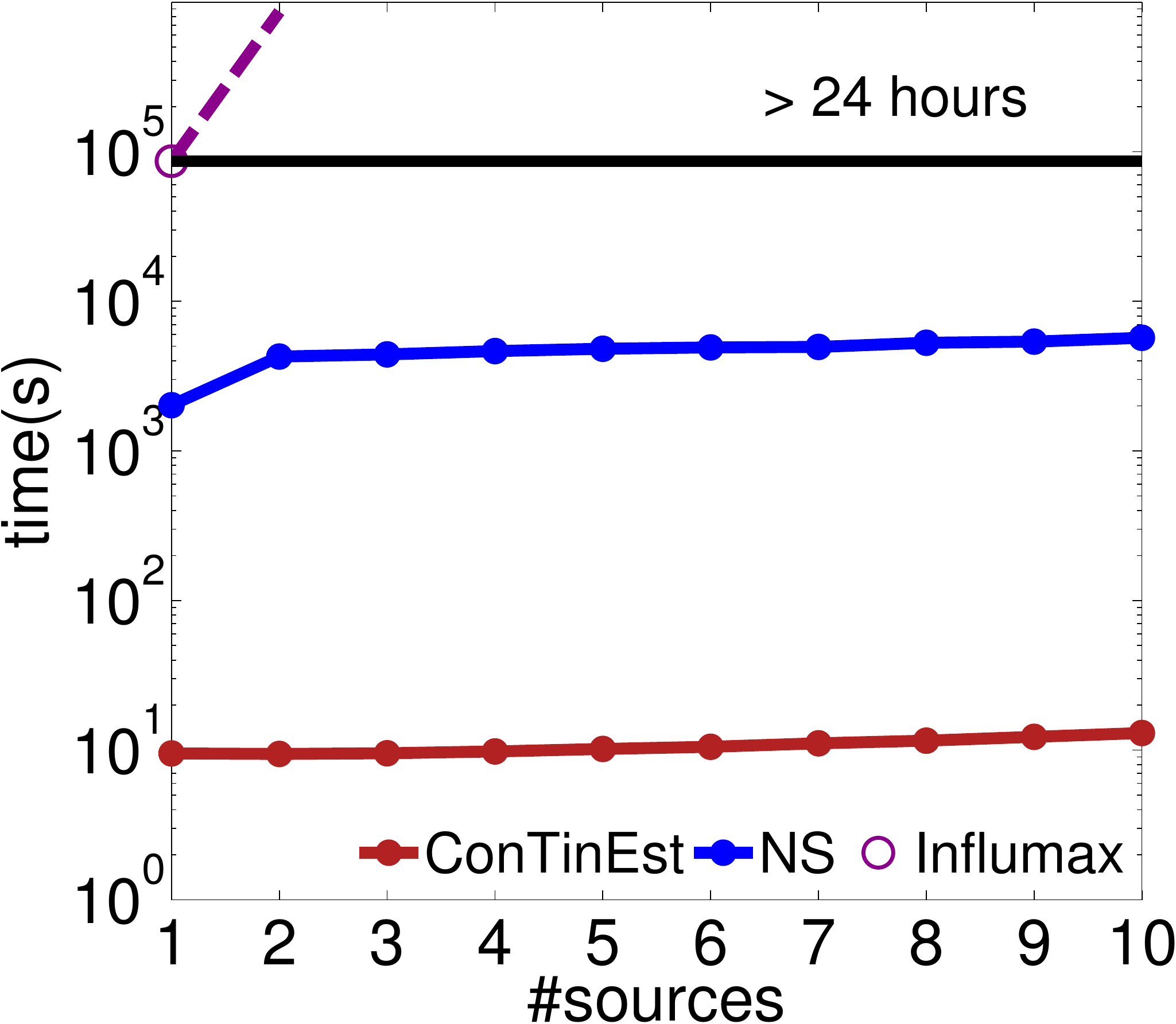} &
\includegraphics[width=0.25\textwidth, height = 95pt]{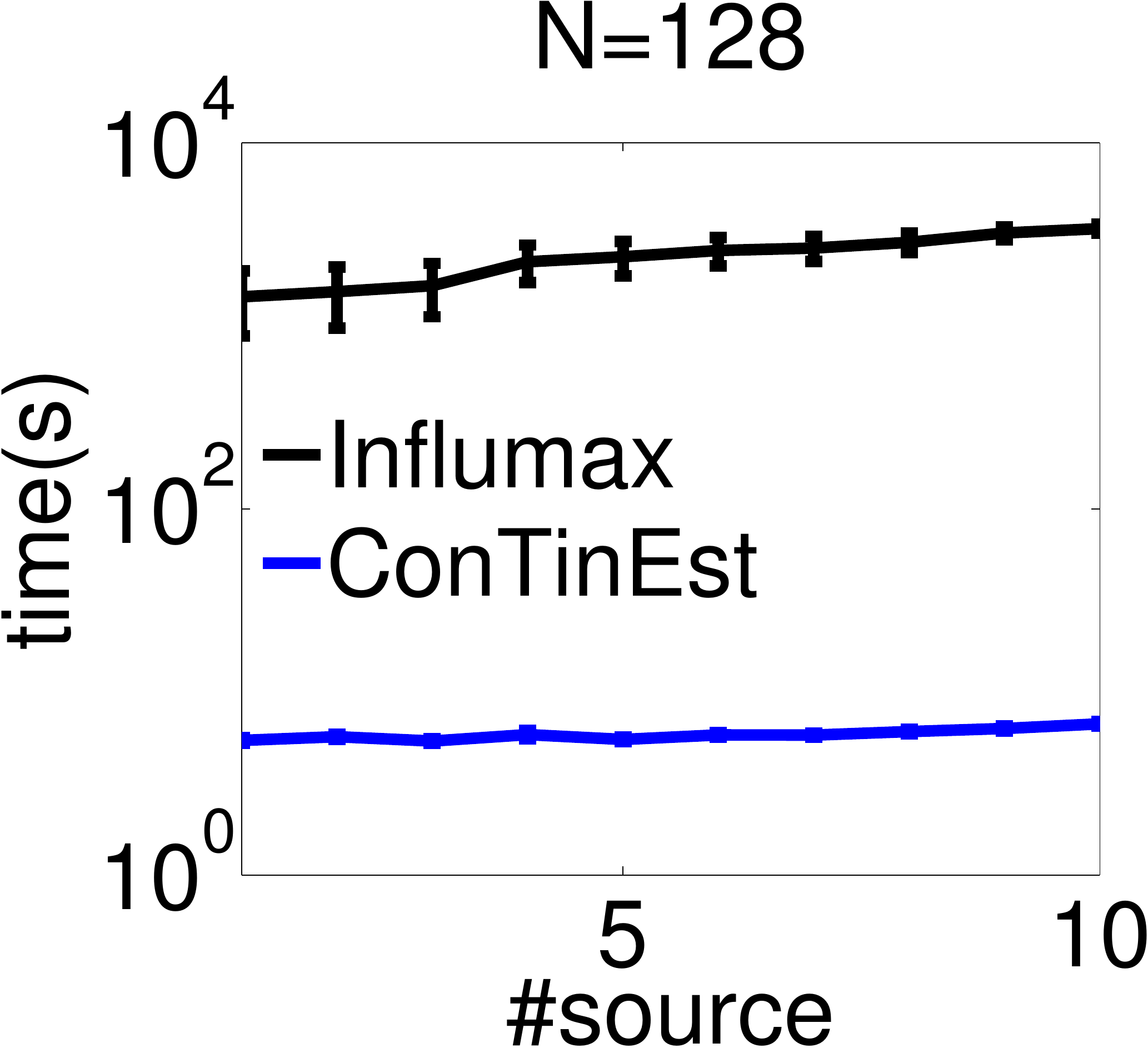} \\
(a) Random & (b)  Hierarchal & (c) Random & (d) Hierarchal
\end{tabular}
 \caption{\label{density_random}Panels(a-b) show the running time against the network density by fixing the number of sources at 10 on the random and hierarchal kronecker network with 128 nodes. Panels(c-d) present the running time as we increase the number of selected source nodes on the networks with 128 nodes and 256 edges. }
\end{figure}
\subsection{Scalability}
\label{app:scalability}
Figure~\ref{density_random} compares \continmax to \influmax and the Naive Simulation (NS) method  in terms of running time for the continuous-time influence maximization problem over the random and hierarchal kronecker type of networks, respectively,  with different densities and sizes on a single 2.4Ghz CPU core. For \continmax, we have drawn 10,000 samples, each of which has 5 random labels assigned to each node. For NS, we follow the work~\cite{kleinberg_kdd03} to run 10,000 Monte Carlo simulations. For running times longer than 24 hours, we use dashed line to qualitatively indicate the estimated performance based on the time complexity of each method.
\end{appendix}

\end{document}